\documentclass[prd,nofootinbib,superscriptaddress,showpacs]{revtex4}
\usepackage[usenames]{color}
\usepackage{amsfonts}
\usepackage{mathtools}
\usepackage{fontenc}
\usepackage{tipa}
\usepackage{wasysym}
\usepackage{tipx}
\usepackage{amsmath}
\usepackage{amssymb}
\usepackage{bm}
\usepackage{url}
\usepackage{dcolumn}
\usepackage{epsfig}
\usepackage{graphicx}
\usepackage{graphics}
\usepackage[latin1]{inputenc}
\usepackage{latexsym}
\usepackage{rotating}
\usepackage{hyperref}
\usepackage{multirow}
\usepackage[caption=false]{subfig}
\usepackage[usenames,dvipsnames]{xcolor}
\usepackage[normalem]{ulem} 
\usepackage{lipsum}
\usepackage[mathscr]{euscript}

\def\ie{{\it i.e.,\ }}

\font\ec=ecrm0800 at 10pt
\def\tho{\hbox{\ec\char'336}}
\def\edt{\hbox{\ec\char'360}}

\usepackage{tipa}

\newcommand{\<}{\begin{equation}}
\newcommand{\?}{\end{equation}}

\newcommand{\kata}{\tho}
\newcommand{\tderivative}{\rho\, \kata^\prime - \rho^\prime \kata }
\newcommand{\rderivative}{\rho\, \kata^\prime + \rho^\prime \kata}
\newcommand{\wacky}{\edt\,\edt^\prime+\edt^\prime\edt}
\newcommand{\bigwacky}{\edt\,\edt^\prime+\edt^\prime\edt - 4\,\rho\,\rho^\prime-4\,\psi_2}

\newcommand{\bea}{\begin{eqnarray}}
\newcommand{\eea}{\end{eqnarray}}
\newcommand{\bsube}{\begin{subequations}}
\newcommand{\esube}{\end{subequations}}
\newcommand{\beq}{\begin{equation}}
\newcommand{\be}{\begin{equation}}
\newcommand{\eeq}{\end{equation}}
\newcommand{\ee}{\end{equation}}

\begin{document}

\title{Gauge-invariant perturbations of Schwarzschild spacetime}

\author{Abhay~G.~Shah}
  
  \affiliation{Mathematical Sciences, University of Southampton, Southampton SO17 1BJ, United Kingdom}
  
  \author{Bernard~F.~Whiting}
  
  \affiliation{Institute for Fundamental Theory, Department of Physics, University of Florida, Gainesville, Florida 32611, USA}

  \author{Steffen~Aksteiner}
  
  \affiliation{Max-Planck-Institut f\"{u}r Gravitationsphysik, Albert-Einstein-Institut, Am M\"{u}hlenberg 1, 14476 Golm, Germany}
  
     \author{Lars~Andersson}
  
 \affiliation{Max-Planck-Institut f\"{u}r Gravitationsphysik, Albert-Einstein-Institut, Am M\"{u}hlenberg 1, 14476 Golm, Germany}
 
    \author{Thomas~B\"{a}ckdahl}   

 \affiliation{Department of Mathematical Sciences, Chalmers University of Technology and the University of Gothenburg, 
SE-412 96 Gothenburg, Sweden}
  
\date{\today}

\begin{abstract}
We study perturbations of Schwarzschild spacetime in a coordinate-free, covariant form. The GHP formulation, having the advantage of not only being covariant but also tetrad-rotation invariant, is used to write down the previously known odd- and even-parity gauge-invariants and the equations they satisfy. Additionally, in the even-parity sector, a new invariant and the second order hyperbolic equation it satisfies are presented. Chandrasekhar's work on transformations of solutions for perturbation equations on Schwarzschild spacetime is translated into the GHP form, \ie solutions for the equations of the even- and odd-parity invariants are written in terms of one another, and the extreme Weyl scalars;  and solutions for the equations of these latter invariants are also written in terms of one another. Recently, further gauge invariants previously used by Steven Detweiler have been described.  His method is translated into GHP form and his \emph{basic} invariants are presented here. We also show how these invariants can be written in terms of curvature invariants. 
\end{abstract}


\maketitle

\section{Overview}
\label{overview}

Whether it is studying gravitational waves, the ring-down phase of comparable mass binary black holes, stability of black holes or the motion of extreme-mass-ratio inspirals, black hole perturbation theory plays a pivotal role.  In this paper we are particularly interested in discussing gauge invariant perturbations which may be identified in the Schwarzschild spacetime, with the hope of developing methods which will eventually be extendible to the Kerr spacetime, and perhaps to most of Petrov type D geometries.  As will be discussed further below, it is known that perturbations of the Weyl scalars $\psi_0$ and $\psi_4$ are spin weighted invariants, while the perturbation of the imaginary part of $\psi_2$ is a scalar invariant.  Given the number of invariants we will consider, we will also attempt to find relations between them, and attempt to specify some minimal set from which all others may be obtained.  We shall explain that the set of scalar invariants which are second order in derivatives of the metric is finite and fully known.  Part of what we do below is explore scalar invariants of higher orders in the derivatives, since it will turn out that, without separation of variables, the Regge-Wheeler and Zerilli invariants are, in fact, scalar and of higher order.  We also find that a set of invariants obtained by Detweiler and others (see \cite{Chen:2016plo}) are also scalar and of higher order in derivatives (when expressed without the separation of variables).  Thus, much of our focus will be on scalar invariants, and we will always consider not using the separation of variables, since we expect that this will not be practicable in the Kerr geometry.  Consequently, we will use GHP notation throughout, since that should eventually allow us to consider most Petrov type D spacetimes without the need to introduce coordinates.

We begin by giving an historical perspective on what has previously been achieved, and how.

\section{Historical Perspective}
\label{intro}

With an interest in exploring the stability of the Schwarzschild black hole, Regge and Wheeler in \cite{Regge-Wheeler} studied linear perturbations or small departures from perfect sphericity of the Schwarzschild spacetime, and this work led them to discover the Regge-Wheeler equation. They started by separating the perturbations into two sectors, even-parity and odd-parity, where parity is governed by the response to the simultaneous transformations, $\theta \rightarrow \pi-\theta$ and $\phi \rightarrow \pi+\phi$. Fields symmetric under this discrete transformation are even parity, fields antisymmetric are odd parity.  The Regge-Wheeler equation is a linear, second-order, hyperbolic equation governing an odd-parity variable. 
Vishveshwara in \cite{Vish} studied the stability of the Schwarzschild metric further, using the Kruskal coordinates. More than a decade later Zerilli, in \cite{Zerilli_PRL}, discovered 
a different linear, second-order hyperbolic equation governing an even-parity variable. 

Subsequently, Moncrief in \cite{Moncrief-1} carried out a detailed study of both of the Regge-Wheeler and Zerilli equations, and wrote the variables which satisfy them in a gauge-invariant form. Cunningham-Price-Moncrief in \cite{CMP} wrote another odd-parity invariant which solved the same Schr\"{o}dinger-type equation as the Regge-Wheeler variable. We now understand that these two odd parity invariants are related by a time derivative modulo the perturbed Einstein tensors, and this has been shown in a covariant manner by Martel and Poisson in \cite{Martel-Poisson}. One can then see that the Cunningham-Price-Moncrief function is proportional to a curvature invariant, $\Im{\dot{\psi}_2}$ (an overdo will represent a perturbed quantity throughout this paper), the imaginary part of the perturbation of the \emph{spin-0} Weyl scalar, $\psi_2$; and the Regge-Wheeler variable is its time derivative (modulo the Einstein tensor components). 

Beginning from a quite different approach, Bardeen and Press in \cite {BP}, on the other hand, studied scalar, electromagnetic and gravitational perturbations of the Schwarzschild black hole, and wrote a master-equation for these perturbations which, subsequently, was generalized to Kerr (and spin-half neutrino fields) by Teukolsky in \cite{Teuk}. In particular, Bardeen and Press studied gravitational test-fields in terms of perturbed \emph{spin-$\pm$2} Weyl scalars, and derived the hyperbolic 
equation they obey. 

Chandrasekhar in \cite{Chandra}, was able to combine all these previous works, having first written the Schr\"{o}dinger-type equations governing the odd- and even-parity invariants as one master equation by showing a relation between the potentials of these two equations; he then related solutions to the equations satisfied by these odd- and even-parity invariants, by writing a solution to one equation in terms of a solution to the other, and its radial derivative. He later went on and calculated the relation between solutions to the equations for the \emph{spin-2} Weyl scalars and solutions to the equations satisfied by the odd- and even-parity invariants, and wrote each one in terms of the other. This allows one to know solutions for one invariant and calculate solutions to the equations for the rest, in any source-free region.  However, we stress that the odd and even parity sectors are completely independent, so Chandrasekhar's approach allowed alternative methods of finding solutions and not a method of relating odd and even parity perturbations for a given source or vacuum spacetime perturbation.

Until now, equations satisfied by the odd- and even-parity gauge-invariant perturbations were written in a coordinate form, whether the usual Boyer-Lindquist or the null Kruskal coordinates. Gerlach and Sengupta in \cite{GS} wrote these equations in a $2+2$ covariant form paving the road to a coordinate-independent study of these perturbations. Sarbach and Tiglio in \cite{ST} took this work further, writing the equations in a completely covariant form, and showed how to get Weyl scalar from these odd- and even-parity invariant amplitudes. Martel and Poisson in \cite{MP} completed these works by providing the gauge-invariant, covariant form of these equations, along with the source terms written in a covariant form, making their work suitable for any coordinate system that one might adopt. All these works were later reviewed by Nagar and Rezzolla in \cite{NR}. Eventually, these perturbations were studied in a light-cone gauge by Preston and Poisson in \cite{PP}. 

Most recently, Aksteiner and Andersson in \cite{AA} wrote the perturbation equations in a Gerald-Held-Penrose (GHP) form and later showed how one can derive the odd- and even-parity equations from a master equation governing perturbations of $\psi_2$.
Steven Detweiler, in his work (of more than a decade ago but only recently described \cite{GI:Det1}), uses an \emph{EZ gauge} which is closely related to the Regge-Wheeler gauge. The metric components in his gauge are directly proportional to the ones in the Regge-Wheeler decomposition up to normalization factors. He finds six basic invariants, four of even-parity and two odd-parity, and writes the Einstein tensors in terms of these invariants. 

In this paper, we present a further chapter in perturbation theory for the Schwarzschild spacetime, hopefully
%
%
paving the way for a new study of perturbations on Kerr spacetime. Throughout, we use the GHP formalism, a coordinate-independent formalism, to write all the inhomogeneous equations with source terms. No separation in terms of radial and angular harmonics is performed. One advantage of using the GHP formalism over the Newman-Penrose (NP) formalism is due to the invariance in terms of a null-tetrad rotation. Another advantage is the obvious prime- and star-symmetries. A prime symmetry, where applying a \emph{prime} implies $\ell^\alpha \leftrightarrow n^\alpha$ and $m^\alpha \leftrightarrow \bar{m}^\alpha$ simultaneously, reduces the number of equations to half the number. A Sachs star symmetry involves $\ell^\alpha \leftrightarrow \pm m^\alpha$ and $n^\alpha \leftrightarrow \pm \bar{m}^\alpha$ ($\pm$ does not refer to ambiguity here; for correct usage see \cite{GHP}). Here ($\ell, n$) form the (out-, in-)going null vector legs of the tetrad, and ($m,\bar{m}$) forms the null-pair on the 2-sphere (where bar implies a complex conjugate). These symmetries can drastically reduce 
the number of independent components of the Einstein tensor one needs to consider for a minimal set. However, see appendix \ref{app:EFE} for a practical set representing the linearized Einstein tensor perturbations.  We write the Einstein field equations, and the known odd- and even-parity equations in GHP form. Einstein field equations and Weyl scalar invariants are then \emph{scalarized} to \emph{GHP type (0,0)} by applying appropriate \emph{angular} operators and/or multiplying with appropriately-weighted Ricci rotation coefficients. 

Among other things, we have discovered a new, second order, hyperbolic differential equation for another even-parity invariant, and this equation is of a lesser formal (angular derivatives included) order than the one solved by the Zerilli invariant. In addition, we have transformed Chandra's work on relating solutions to the equations satisfied by the odd- and even-parity invariants, and writing the solutions to one in terms of solutions to the other in a coordinate-independent, 
GHP form. Their relation with solutions to the equations satisfied by the extreme Weyl scalars is also shown in GHP form. Following Steven Detweiler's work, we calculate his invariants in GHP form and show how they are related. One question that we try to address here is whether all known invariants are related to each other by curvature invariants and their derivatives (perturbed Einstein tensor components, $\dot{\psi}_0$, $\dot{\psi}_4$, and $\Im{\dot{\psi}_2}$). And, we see that it is possible to do so. 

We organize our work in the following manner. In the next section, Sec. \ref{sec:ghp}, we summarize the NP and GHP formalisms. We refer the interested reader to the original works in \cite{NP, GHP} for further explanation, and only present details relevant to the current work. In Sec. \ref{sec:gauge-invariants}, we present the already known odd- and even-parity invariants, along with the equations they obey (including the source terms) in a completely covariant, unseparated form, which involves reinstating many suppressed angular derivatives. We also show how a small transformation leads us to write the two odd-parity equations as one. In Sec. \ref{FCA_1}, we show how solutions to the equations for the even and odd parity invariants can be written in terms of each other. We also show how to write either in terms of solutions to the equations for $\dot{\psi}_0$ or $\dot{\psi}_4$, and vice versa. For these relations to be possible, being in a background, source-free region is a necessary condition. In the following section, Sec. \ref{sec:recipe}, we translate Steven Detweiler's work into an unseparated, GHP form, and show how to obtain further invariants. In principle, it is possible to get as many invariants as one wishes, but they will all be of order higher than for these basic invariants. We then show how these \emph{basic} invariants can be written in terms of curvature invariants. In the last section, Sec. \ref{sec:concl}, we summarize our work, leave the reader with some additional remarks, and indicate where we wish to take this work in future. 
In App.\ \ref{metric_comp}, we show the relation between the GHP components of metric perturbation and those of tensor harmonics. In App.\ \ref{app:EFE}, we write the scalarized, GHP type (0,0) form of perturbed Einstein tensors, and in App.\ \ref{app:coord}, we show how to translate GHP quantities to a coordinate form and present the new even-parity equation in coordinate form. The signature used throughout this work is -2, \ie $(+---)$.

\section{NP and GHP formulation}
\label{sec:ghp}

Newman and Penrose in \cite{NP} used a novel approach to study the general theory of relativity by introducing a complex, null tetrad ($\ell^\alpha$, $n^\alpha$, $m^\alpha$, $\bar{m}^\alpha$), and all the known equations were projected on this tetrad. In this work, for simplicity, we will enumerate the tetrad as ($e_1^\alpha=\ell^\alpha$, $e_2^\alpha=n^\alpha$, $e_3^\alpha=m^\alpha$, $e_4^\alpha=\bar{m}^\alpha$). The orthogonality relation between them is
\begin{eqnarray}
e_1\cdot e_2=1, \quad e_3\cdot e_4=-1,
\end{eqnarray}
with rest of the dot-products being zero. The background metric is then written as 
\begin{eqnarray}
g_{\alpha\beta} = \ell_\alpha n_\beta + n_\alpha \ell_\beta - m_\alpha \bar{m}_\beta - \bar{m}_\alpha m_\beta.
\end{eqnarray}
The derivative operators along ($\ell^\alpha$, $n^\alpha$, $m^\alpha$, $\bar{m}^\alpha$) are ($D$, $\Delta$, $\delta$, $\bar{\delta}$), and for the case at hand (Schwarzschild background), the non-zero spin coefficients are $\alpha$, $\beta$, $\rho$, $\epsilon$, $\gamma$, and $\mu$. We refer the readers to \cite{NP} for further details. 

Under a tetrad transformation 
\begin{eqnarray} \label{null-transformation}
\ell^\alpha \rightarrow \lambda \,\bar{\lambda}\, \ell^\alpha&,& \quad n^\alpha \rightarrow \frac{1}{\lambda \,\bar{\lambda}} n^\alpha, \nonumber \\
m^\alpha \rightarrow \frac{\lambda}{ \bar{\lambda}} m^\alpha&,& \quad \bar{m}^\alpha \rightarrow \frac{ \bar{\lambda}}{\lambda} \bar{m}^\alpha,
\end{eqnarray}
the metric and the orthogonality relations above are preserved. Regarding such a transformation, if a scalar, $\Psi$, transforms as
\begin{eqnarray}
\Psi \rightarrow \lambda^p \, \bar{\lambda}^q \, \Psi,
\end{eqnarray}
it is said to be of \emph{type, (p,q),} where the \emph{spin-weight} is $(p-q)/2$, and the \emph{boost-weight} is $(p+q)/2$. As an example, the Weyl scalar, $\psi_0$ ($=-C_{\alpha\beta\gamma\delta}\ell^\alpha m^\beta\ell^\gamma m^\delta$), is of type (4,0), has \emph{spin-weight} 2, and \emph{boost-weight} 2.  An important symmetry is the prime-symmetry in the GHP formalism, where the \emph{prime-operator} implies
\begin{eqnarray}
(\ell^\alpha)^\prime = n^\alpha&,& \quad (n^\alpha)^\prime = \ell^\alpha, \nonumber \\
(m^\alpha)^\prime = \bar{m}^\alpha&,& \quad (\bar{m}^\alpha)^\prime = m^\alpha,
\end{eqnarray}
where
\begin{eqnarray}
&\rho^\prime = -\mu, \quad \epsilon^\prime = -\gamma, \quad \beta^\prime = -\alpha.
\end{eqnarray}
The \emph{prime-operator} changes a scalar of type $(p,q)$ to type $(-p,-q)$, and complex conjugation changes it to type $(q,p)$.

We now introduce GHP derivative operators (\Thorn, $\edt$), and their primes. When acting on a scalar, $\Psi$,  of type $(p,q)$, these operators are
\begin{eqnarray}
\kata \,\Psi &=& (D-p\epsilon-q\bar{\epsilon})\Psi, \nonumber \\ 
\kata^\prime \,\Psi &=& (D^\prime+p\epsilon^\prime+q\bar{\epsilon}^\prime)\Psi, \nonumber \\
\edt \,\Psi &=& (\delta-p\beta+q\bar{\beta}^\prime)\Psi, \,\text{and}\nonumber \\
\edt^\prime \,\Psi &=& (\delta^\prime+p\beta^\prime-q\bar{\beta}^\prime) \Psi.
\end{eqnarray}
This combination guarantees no derivatives of the parameter $\lambda$ under the transformation eq.\ \eqref{null-transformation}.

Finally, we present the relevant Ricci identities, 
\begin{eqnarray}
\kata \, \rho &=& \rho^2, \nonumber \\
\kata \, \rho^\prime &=& \rho \, \rho^\prime - \psi_2, \nonumber \\
\kata \,\psi_2 &=& 3\,\rho\,\psi_2
\end{eqnarray}
Their \emph{primes} also hold.


\section{Gauge-invariants}
\label{sec:gauge-invariants}
To construct some basic gauge invariants, we first present the metric perturbation in GHP form, letting
\begin{align}
h_{ab}=&h_{22}\left(\ell_a\ell_b\right)+h_{11}\left(n_a n_b\right)+h_{12}\left( \ell_a n_b + n_a \ell_b\right) + h_{33}\left(\bar{m}_a\bar{m}_b\right) + h_{44}\left(m_a m_b\right) + h_{34} (m_a \bar{m}_b + \bar{m}_a m_b) \nonumber \\
&-h_{24}\left( \ell_a m_b + m_a \ell_b\right)-h_{23}\left( \ell_a \bar{m}_b + \bar{m}_a \ell_b\right) -h_{14}\left( n_a m_b + m_a n_b\right)-h_{13}\left( n_a \bar{m}_b + \bar{m}_a n_b\right).
\end{align}

We now write the transformation of the metric perturbation under a linearized diffeomorphism, $(\delta_\xi h)_{ab}=\nabla_{(a}\xi_{b)}$, in GHP form,
\begin{align} \label{gaugetransf}
\allowdisplaybreaks
(\delta_\xi h)_{11} &= 2\,\kata\,\xi_1, \nonumber \\
(\delta_\xi h)_{12} &= \kata^\prime \,\xi_1 + \kata \,\xi_2, \nonumber \\ 
(\delta_\xi h)_{22} &= 2\,\kata^\prime\,\xi_2, \nonumber \\ 
(\delta_\xi h)_{13} &= \edt\,\xi_1 + (\kata+\rho)\,\xi_3, \nonumber \\
(\delta_\xi h)_{14} &= \edt^\prime\,\xi_1 + (\kata+\rho)\,\xi_4, \nonumber \\
(\delta_\xi h)_{23} &= \edt\,\xi_2 + (\kata^\prime+\rho^\prime)\,\xi_3, \nonumber \\
(\delta_\xi h)_{24} &= \edt^\prime\,\xi_2 + (\kata^\prime+\rho^\prime)\,\xi_4, \nonumber \\
(\delta_\xi h)_{33} &= 2\,\edt\,\xi_3, \nonumber \\
(\delta_\xi h)_{44} &= 2\,\edt^\prime\,\xi_4, \nonumber \\
(\delta_\xi h)_{34} &= 2\,\rho^\prime\,\xi_1+ 2\,\rho\,\xi_2 +\edt^\prime\,\xi_3 + \edt\,\xi_4,
\end{align}
where $h_{ij}$ is the projection of the metric perturbation on the null tetrad, and $\xi_i$ is the gauge vector projected on the null tetrad. At any stage these can be plugged in to check the gauge-invariance of the invariants found in this work. 

The gauge-invariants we construct, or translate in the GHP form from previous works, are all of type $(0,0)$. This has meant constructing them from the following basic set
\begin{eqnarray}
\rho^{\prime\,2}\,h_{11}, \,\,\, \rho^2\,h_{22},\,\,\, \edt^\prime\,\edt^\prime\,h_{33}, \,\,\, \edt\,\edt\,h_{44}, \,\,\, h_{12}, \nonumber \\
\rho^\prime \, \edt^\prime \, h_{13}, \,\,\, \rho^\prime \, \edt \, h_{14}, \,\,\, \rho \, \edt^\prime \, h_{23}, \,\,\, \rho \, \edt \, h_{24}, \,\,\, h_{34}.
\end{eqnarray}
Odd-parity gauge-invariants are constructed by subtracting a quantity from its complex conjugate. This implies that the odd-parity gauge-invariants can only be constructed from linear combinations of $h_{13}$, $h_{14}$, $h_{23}$, $h_{24}$, $h_{33}$, and $h_{44}$. On the other hand, even parity gauge-invariants can be constructed from any of these ten components. In this work, we will be presenting our gauge-invariants in terms of these basic type (0,0) quantities, which are easy to relate to the Regge-Wheeler-Zerilli metric components as shown in Appendix A.  The three invariants and the equations presented in this section are also invariant under the \emph{prime}-symmetry (up to a possible minus sign for the invariants themselves).

\subsection{Odd Parity Invariants}\label{OPI_1}
We now present the first odd-parity invariant in GHP form, the Cunningham-Price-Moncrief function,
\begin{widetext}
\begin{align}\label{IO1}
I_{O_{1}} =& \big( \rho\,\kata^\prime - \rho^\prime\,\kata \big)\big( \rho^\prime \, \edt^\prime \, h_{13} - \rho^\prime \, \edt \, h_{14} + \rho \, \edt^\prime \, h_{23} - \rho\,\edt\,h_{24}\big) \nonumber \\
&+\big( \rho\,\kata^\prime + \rho^\prime\,\kata -2\,\rho\,\rho^\prime\big)\big( \rho^\prime \, \edt^\prime \, h_{13} - \rho^\prime \, \edt \, h_{14} - \rho \, \edt^\prime \, h_{23} + \rho\,\edt\,h_{24}\big)
\equiv-2\rho\rho'\Im{\dot{\psi}_2}
\end{align}
\end{widetext}
which is proportional to $\Im{\dot{\psi}_2}$. A related invariant can also be presented in the following form,
\begin{eqnarray}\label{IO1a}
I_{O_{1a}} = \kata\, \edt^\prime \,h_{23} - \kata^\prime \, \edt^\prime \, h_{13} - \kata \, \edt \, h_{24} + \kata^\prime\,\edt \, h_{14}\equiv\Im{\dot{\psi}_2}.
\end{eqnarray}
which is exactly $\Im{\dot{\psi}_2}$. The relation between the two forms of this invariant is $I_{O_{1}}=-2\rho\rho^\prime I_{O_{1a}}$. We bring the reader's attention to the operators $\left( \rho\,\kata^\prime - \rho^\prime\,\kata \right)$ and $\left( \rho\,\kata^\prime + \rho^\prime\,\kata \right)$ in $I_{O_1}$. Using the Kinnersley tetrad, these operators are directly proportional to the time- and radial- derivatives, respectively, and become
\begin{eqnarray}
\left( \rho\,\kata^\prime - \rho^\prime\,\kata \right) &=& \frac{-1}{r} \partial_t\nonumber \\
\left( \rho\,\kata^\prime + \rho^\prime\,\kata \right) &=& \frac{1}{r} \partial_{r_*}
\end{eqnarray}
where $r_*$ is the tortoise-coordinate.  The equation satisfied by $I_{O_{1a}}$ is 
\begin{widetext}
\begin{eqnarray}
&&\Big[\big( \rho\,\kata^\prime + \rho^\prime\,\kata \big)\big( \rho\,\kata^\prime + \rho^\prime\,\kata \big) - \big( \rho\,\kata^\prime - \rho^\prime\,\kata \big) \big( \rho\,\kata^\prime - \rho^\prime\,\kata \big) - 14\rho\rho^\prime \big( \rho\,\kata^\prime + \rho^\prime\,\kata \big) - 2\rho \rho^\prime \big( \edt\edt^\prime + \edt^\prime\edt-12\rho\rho^\prime-12\psi_2\big) \Big]I_{O_{1a}} \nonumber \\
&&=2\big( \rho\,\kata^\prime + \rho^\prime\,\kata - 6\rho\rho^\prime\big)\big( \rho^\prime \, \edt^\prime \, \mathcal{E}_{13} - \rho^\prime \, \edt \, \mathcal{E}_{14} - \rho \, \edt^\prime \, \mathcal{E}_{23} + \rho\,\edt\,\mathcal{E}_{24}\big) \nonumber \\
&&+2\big( \rho\,\kata^\prime - \rho^\prime\,\kata\big)\big( \rho^\prime \, \edt^\prime \, \mathcal{E}_{13} - \rho^\prime \, \edt \, \mathcal{E}_{14} + \rho \, \edt^\prime \, \mathcal{E}_{23} - \rho\,\edt\,\mathcal{E}_{24}\big)
\end{eqnarray}
\end{widetext}
where $\mathcal{E}_{ab}$ is the GHP-projected, linearized Einstein tensor (see also App.\ \ref{app:EFE}).  We also present the source in terms of the basic type (0,0) quantities where derivatives of $h$ are replaced by $\mathcal{E}$, as these are then easy to convert into the Regge-Wheeler-Zerilli source terms (again, see App.\ \ref{app:coord}). 

Another odd-parity invariant, related to $I_{O_{1a}}$ (loosely, by a time-derivative), is the Regge-Wheeler invariant,
\begin{widetext}
\begin{eqnarray}\label{IO2}
I_{O_2} &=& \big(\edt\, \edt^\prime + \edt^\prime\edt-4\,\rho\,\rho^\prime-4\,\psi_2\big)\big(\rho^\prime\edt^\prime h_{13}-\rho^\prime\edt\, h_{14}+\rho\,\edt^\prime h_{23}-\rho\,\edt\,h_{24}\big) \nonumber \\ 
&&- \big(\rho^\prime\kata+\rho\,\kata^\prime-4\,\rho\,\rho^\prime\big)\big(\edt^\prime\edt^\prime h_{33}-\edt\,\edt\, h_{44}\big).
\end{eqnarray}
\end{widetext}
The equation it satisfies is
\begin{widetext}
\begin{eqnarray}\label{RW}
&&\Big[\big( \rho\,\kata^\prime + \rho^\prime\,\kata \big)\big( \rho\,\kata^\prime + \rho^\prime\,\kata \big) - \big( \rho\,\kata^\prime - \rho^\prime\,\kata \big) \big( \rho\,\kata^\prime - \rho^\prime\,\kata \big) - 18\rho\rho^\prime \big( \rho\,\kata^\prime + \rho^\prime\,\kata \big) - 2\rho \rho^\prime \big( \edt\edt^\prime + \edt^\prime\edt-24\rho\rho^\prime-14\psi_2\big) \Big]I_{O_{2}} \nonumber \\
&&=2\big( \rho\,\kata^\prime + \rho^\prime\,\kata - 8\rho\rho^\prime\big) \big( \rho\,\kata^\prime - \rho^\prime\,\kata \big) \big( \rho^\prime \, \edt^\prime \, \mathcal{E}_{13} - \rho^\prime \, \edt \, \mathcal{E}_{14} - \rho \, \edt^\prime \, \mathcal{E}_{23} + \rho\,\edt\,\mathcal{E}_{24}\big) \nonumber \\
&&\,\,\,\,\,+2\Big[ \big( \rho\,\kata^\prime + \rho^\prime\,\kata \big) \big( \rho\,\kata^\prime + \rho^\prime\kata \big) - 18\rho\,\rho^\prime \big( \rho\,\kata^\prime + \rho^\prime\,\kata\big) - 2\rho \rho^\prime \big( \edt\edt^\prime + \edt^\prime\edt-24\rho\rho^\prime-14\psi_2\big) \Big]\nonumber \\
&& \qquad\times\big( \rho^\prime \, \edt^\prime \, \mathcal{E}_{13} - \rho^\prime \, \edt \, \mathcal{E}_{14} + \rho \, \edt^\prime \, \mathcal{E}_{23} - \rho\,\edt\,\mathcal{E}_{24}\big).
\end{eqnarray}
\end{widetext}
The exact relation between the Cunningham-Price-Moncrief invariant and the Regge-Wheeler invariant is
\begin{widetext}
\begin{align}\label{CPM-RW}
\big(\rho\,\kata^\prime - \rho^\prime \kata\big)\,I_{O_1} = -2\,\rho\,\rho^\prime I_{O_2} + 4\,\rho\,\rho^\prime \big( \rho^\prime\edt^\prime \mathcal{E}_{13} - \rho^\prime\edt\,\mathcal{E}_{14}+\rho\,\edt^\prime \mathcal{E}_{23} - \rho\,\edt\,\mathcal{E}_{24} \big).
\end{align}
\end{widetext}
It is relatively straightforward to check these relations, and the validity of the equations presented, by brute-force, simply by substitution using the Einstein tensor components presented in Appendix B. We would like to bring the reader's attention to the potentials involved in the two odd-parity equations above. They appear to be different when compared to the covariant Eqs (5.14) and (5.19) of Martel-Poisson \cite{MP} which have the same potential for both of them. This difference can be removed by re-scaling the two invariants, $I_{O_{1a}}$ and $I_{O_2}$ appropriately, which will be shown later.

With this we now move to the even-parity sector. 

\subsection{Even Parity Invariants}\label{EPI_1}
As is well known, the Zerilli potential involves division by an algebraic factor which involves both $r$ and $\ell$, the angular eigenvalue.  Thus to write the equation without division by the angular operator, many extra angular factors need to be used in both the Zerilli equation and, in fact, in the Zerilli variable itself too.  We find Zerilli's even-parity invariant in GHP form is
\begin{widetext}
\begin{eqnarray}\label{eq:zervar}
I_Z &=& \big( \edt\edt^\prime + \edt^\prime\edt-4\,\rho\,\rho^\prime-4\,\psi_2\big)\big( \edt\edt^\prime + \edt^\prime\edt\big) \big( \rho\,\kata^\prime - \rho^\prime\,\kata \big) h_{34} \nonumber \\
&& +\big( \edt\edt^\prime + \edt^\prime\edt-4\,\rho\,\rho^\prime-4\,\psi_2\big)\big( \edt\edt^\prime + \edt^\prime\edt\big) \big( \rho^{\prime 2} h_{11} - \rho^2 h_{22}  \big) \nonumber \\
&& + \big( \edt\edt^\prime + \edt^\prime\edt-4\,\rho\,\rho^\prime+2\,\psi_2\big) \big( \rho\,\kata^\prime - \rho^\prime\,\kata \big) \big(  \edt^\prime \edt^\prime h_{33} + \edt\,\edt\,h_{44}  \big) \nonumber \\
&& + \big( \edt\edt^\prime + \edt^\prime\edt-4\,\rho\,\rho^\prime-4\,\psi_2\big) \big( \rho\,\kata^\prime + \rho^\prime\,\kata -6\,\rho\,\rho^\prime + 2 \psi_2 \big) \big( \rho^\prime \edt^\prime h_{13} + \rho^\prime \edt\, h_{14} - \rho\,\edt^\prime h_{23} - \rho\,\edt\,h_{24} \big) \nonumber \\
&& + \big( \edt\edt^\prime + \edt^\prime\edt-4\,\rho\,\rho^\prime-4\,\psi_2\big) \big( \rho\,\kata^\prime - \rho^\prime\,\kata \big) \big( \rho^\prime \edt^\prime h_{13} + \rho^\prime \edt\, h_{14} + \rho\,\edt^\prime h_{23} + \rho\,\edt\,h_{24} \big),
\end{eqnarray}
\end{widetext}
and the equation it satisfies is given by
\begin{widetext}
\begin{eqnarray}\label{eq:zereq}
&&\big( \hat{\mathcal{X}}-4\,\rho\,\rho^\prime+2\,\psi_2\big)^2 \big( \rho\,\kata^\prime + \rho^\prime\,\kata \big) \big( \rho\,\kata^\prime + \rho^\prime\,\kata \big) I_{Z} \nonumber \\
&-&\big( \hat{\mathcal{X}}-4\,\rho\,\rho^\prime+2\,\psi_2\big)^2 \big( \rho\,\kata^\prime - \rho^\prime\,\kata \big) \big( \rho\,\kata^\prime - \rho^\prime\,\kata \big) I_{Z} \nonumber \\
&-& 2\,\rho\,\rho^\prime \big( \hat{\mathcal{X}}-4\,\rho\,\rho^\prime+2\,\psi_2\big) \big( 13\,\hat{\mathcal{X}}-52\,\rho\,\rho^\prime+38\,\psi_2\big) \big( \rho\,\kata^\prime + \rho^\prime\,\kata \big) I_Z \nonumber \\
&-&2\,\rho\,\rho^\prime \big( \hat{\mathcal{X}}-4\,\rho\,\rho^\prime+2\,\psi_2\big) \Big[ \hat{\mathcal{X}}^2-16\big(4\,\rho\,\rho^\prime+\psi_2\big)\hat{\mathcal{X}} -24\big( \psi_2-\rho\,\rho^\prime \big)  \big( \psi_2+10\,\rho\,\rho^\prime \big)\Big] I_Z\nonumber \\
= \nonumber \\
&&4\,\rho\,\rho^\prime \big( \hat{\mathcal{X}} - 4\,\rho\,\rho^\prime-4\,\psi_2 \big) \big( \hat{\mathcal{X}} - 4\,\rho\,\rho^\prime+2\,\psi_2 \big)^2 \big( \rho\,\kata^\prime + \rho^\prime\,\kata \big) \big(\rho^\prime \edt^\prime \mathcal{E}_{13} + \rho^\prime \edt\, \mathcal{E}_{14} - \rho\,\edt^\prime \mathcal{E}_{23} - \rho\,\edt\, \mathcal{E}_{24} \big) \nonumber \\
&-& \hat{\mathcal{X}} \big( \hat{\mathcal{X}} - 4\,\rho\,\rho^\prime-4\,\psi_2 \big) \big( \hat{\mathcal{X}} - 4\,\rho\,\rho^\prime+2\,\psi_2 \big)^2 \big( \rho\,\kata^\prime + \rho^\prime\,\kata \big) \big(\rho^{\prime 2} \mathcal{E}_{11} - \rho^2, \mathcal{E}_{22} \big) \nonumber \\
&-& \hat{\mathcal{X}} \big( \hat{\mathcal{X}} - 4\,\rho\,\rho^\prime-4\,\psi_2 \big) \big( \hat{\mathcal{X}} - 4\,\rho\,\rho^\prime+2\,\psi_2 \big)^2 \big( \rho\,\kata^\prime - \rho^\prime\,\kata \big) \big(\rho^{\prime 2} \mathcal{E}_{11} + 2\,\rho\,\rho^\prime \mathcal{E}_{12} +  \rho^2, \mathcal{E}_{22}\big) \nonumber \\
&+& 4\,\rho\,\rho^\prime \big( \hat{\mathcal{X}} - 4\,\rho\,\rho^\prime+2\,\psi_2 \big)^3 \big( \rho\,\kata^\prime - \rho^\prime\,\kata \big) \big( \edt^\prime\edt^\prime \mathcal{E}_{33} + \edt\,\edt\, \mathcal{E}_{44} \big) \nonumber \\
&+&4\,\rho\,\rho^\prime \big( \hat{\mathcal{X}} - 4\,\rho\,\rho^\prime-4\,\psi_2 \big) \big( \hat{\mathcal{X}} - 4\,\rho\,\rho^\prime+2\,\psi_2 \big)^2 \big( \rho\,\kata^\prime - \rho^\prime\,\kata \big) \big(\rho^\prime \edt^\prime \mathcal{E}_{13} + \rho^\prime \edt\, \mathcal{E}_{14} + \rho\,\edt^\prime \mathcal{E}_{23} + \rho\,\edt\, \mathcal{E}_{24} \big) \nonumber \\
&+&4\,\rho\,\rho^\prime \hat{\mathcal{X}} \big( \hat{\mathcal{X}} - 4\,\rho\,\rho^\prime+2\,\psi_2 \big) \big( \hat{\mathcal{X}} - 4\,\rho\,\rho^\prime+8\,\psi_2 \big) \big( \hat{\mathcal{X}} - 4\,\rho\,\rho^\prime-4\,\psi_2 \big) \big(\rho^{\prime 2}  \mathcal{E}_{11} - \rho^2 \mathcal{E}_{22} \big) \nonumber \\
&+&2\,\rho\,\rho^\prime \big( \hat{\mathcal{X}} - 4\,\rho\,\rho^\prime-4\,\psi_2 \big) \big( \hat{\mathcal{X}} - 4\,\rho\,\rho^\prime+2\,\psi_2 \big) \Big[ \hat{\mathcal{X}}^2 -2\,\hat{\mathcal{X}} \big( \psi_2 + 8\,\rho\,\rho^\prime \big) - 8\big( \psi_2^2 + 7\,\rho\,\rho^\prime\psi_2-6,\rho^2\rho^{\prime 2} \big)  \Big] \nonumber \\
&& \times \big(\rho^\prime \edt^\prime \mathcal{E}_{13} + \rho^\prime \edt\, \mathcal{E}_{14} - \rho\,\edt^\prime \mathcal{E}_{23} - \rho\,\edt\, \mathcal{E}_{24} \big),
\end{eqnarray}
\end{widetext}
where $\hat{\mathcal{X}} = \big( \edt\,\edt^\prime + \edt^\prime\edt \big)$ is used for brevity.  
Once again, one can straightforwardly check the validity of these equations by using App.\ \ref{app:EFE}. 

\subsubsection{A New Gauge Invariant}\label{NGI_1}
Finally in this section, we present a new even-parity gauge-invariant which also obeys a $2^{nd}$ order hyperbolic equation. The invariant is 
\begin{widetext}
\begin{align}
{I}_{\nu E1} =& \big( \hat{\mathcal{X}}-4\,\rho\,\rho^\prime-4\,\psi_2 \big)\bigg[ \rho^{\prime 2}h_{11} + 2\,\rho\,\rho^\prime h_{12} + \rho^2 h_{22}  + \left( 2\,\rho\,\rho^\prime-\rho^\prime\kata - \rho\,\kata^\prime -\frac12 \hat{\mathcal{X}}-\psi_2\right)\,h_{34} \nonumber \\
&+\rho^\prime \edt^\prime h_{13} + \rho^\prime \edt\,h_{14} + \rho\,\edt^\prime h_{23} + \rho\,\edt\,h_{24} + \frac12 \edt^\prime\edt^\prime h_{33} + \frac12  \edt\,\edt\,h_{44}  \bigg] + 3\,\psi_2 \big( \edt^\prime\edt^\prime h_{33} + \edt\,\edt\,h_{44} \big),
\end{align}
\end{widetext}
which is of a total differential order (4), less than that for the Zerilli variable (5 --- see equation  eq.\ \eqref{eq:zervar}).  
The equation our new invariant obeys is given by
\begin{widetext}
\begin{align} \label{new_eqn}
&2\,\big(\hat{\mathcal{X}}-4\,\rho\,\rho^\prime+2\,\psi_2\big) \Big[\big( \rho\,\kata^\prime + \rho^\prime\kata \big)^2-\big( \rho\,\kata^\prime - \rho^\prime\kata \big)^2\Big] \, {I}_{\nu E1} -4\,\rho\,\rho^\prime\big( 11\,\hat{\mathcal{X}} -44\,\rho\,\rho^\prime+34\,\psi_2\big) \big( \rho\,\kata^\prime + \rho^\prime\kata \big)\, {I}_{\nu E1} \nonumber \\
+&\,4\,\rho\,\rho^\prime \Big[ -\hat{\mathcal{X}}^2+2\big(  7\,\psi_2+22\,\rho\,\rho^\prime \big)\,\hat{\mathcal{X}}+20\,\big( \psi_2^2+8\,\rho\,\rho^\prime-8\,\rho^2\rho^{\prime 2} \big) \Big]\,{I}_{\nu E1} \nonumber \\
&=  \big(\hat{\mathcal{X}}-4\,\rho\,\rho^\prime+2\,\psi_2\big)\, \big(\hat{\mathcal{X}}-4\,\rho\,\rho^\prime-4\,\psi_2\big)\, \hat{\mathcal{X}} \, \big(  \rho^{\prime 2}\mathcal{E}_{11} + 2\,\rho\,\rho^\prime \mathcal{E}_{12} + \rho^2 \mathcal{E}_{22}  \big) \nonumber \\
-&\,4\,\rho\,\rho^\prime\big(\hat{\mathcal{X}}-4\,\rho\,\rho^\prime+2\,\psi_2\big)\, \big(\hat{\mathcal{X}}-4\,\rho\,\rho^\prime+2\,\psi_2\big)\,\big( \edt^\prime\edt^\prime \mathcal{E}_{33} + \edt\,\edt\,\mathcal{E}_{44} \big) \nonumber \\
-&\,2\,\big(\hat{\mathcal{X}}-4\,\rho\,\rho^\prime+2\,\psi_2\big)\, \big(\hat{\mathcal{X}}-4\,\rho\,\rho^\prime-4\,\psi_2\big)\,\big( \rho\,\kata^\prime - \rho^\prime\kata \big)\,\big(  \rho^{\prime 2}\mathcal{E}_{11} - \rho^2 \mathcal{E}_{22}  \big) \nonumber \\
-&\,4\,\rho\,\rho^\prime\big(\hat{\mathcal{X}}-4\,\rho\,\rho^\prime+2\,\psi_2\big)\, \big(\hat{\mathcal{X}}-4\,\rho\,\rho^\prime-4\,\psi_2\big)\, \big( \rho^\prime \edt^\prime \mathcal{E}_{13} + \rho^\prime \edt\,\mathcal{E}_{14} + \rho\,\edt^\prime \mathcal{E}_{23} + \rho\,\edt\,\mathcal{E}_{24} \big) \nonumber \\
-&\,2\,\big(\hat{\mathcal{X}}-4\,\rho\,\rho^\prime+2\,\psi_2\big)\, \big(\hat{\mathcal{X}}-4\,\rho\,\rho^\prime-4\,\psi_2\big)\,\big( \rho\,\kata^\prime + \rho^\prime\kata \big)\, \big(  \rho^{\prime 2}\mathcal{E}_{11} - 2\,\rho\,\rho^\prime \mathcal{E}_{12} + \rho^2 \mathcal{E}_{22}  \big) \nonumber \\
-&\, \Big[ \hat{\mathcal{X}}^2 + \hat{\mathcal{X}}\,\big( 2\,\psi_2-20\,\rho\,\rho^\prime \big) + 64\,\rho^2\rho^{\prime 2} - 80\,\rho\,\rho^\prime\psi_2 \Big] \, \big(\hat{\mathcal{X}}-4\,\rho\,\rho^\prime-4\,\psi_2\big)\, \big(  \rho^{\prime 2}\mathcal{E}_{11} - 2\,\rho\,\rho^\prime \mathcal{E}_{12} + \rho^2 \mathcal{E}_{22}  \big).
\end{align}
\end{widetext}
Compared to Zerilli's $6^{th}$ order equation \eqref{eq:zereq}, this equation is of $4^{th}$ order. Moreover, the Zerilli invariant is $5^{th}$ order in derivatives acting on the metric perturbation and the one above is of $4^{th}$ order. Effectively, the Zerilli's equation is $11^{th}$ order acting on the metric perturbation whereas the one above is 3 orders less, i.e., $8^{th}$ order on the metric perturbation.

One would have noticed that two expressions, $\big(\hat{\mathcal{X}}-4\,\rho\,\rho^\prime+2\,\psi_2\big)$ and $\big(\hat{\mathcal{X}}-4\,\rho\,\rho^\prime-4\,\psi_2\big)$, occur frequently in these calculations. Using Kinnersley tetrad, these correspond to $-(\ell-1)(\ell+2)/r^2$ and $-(\ell-1)(\ell+2)/r^2\,-\,6M/r^3$, respectively, in their separated forms. And $\hat{\mathcal{X}}$ by itself is $-\ell(\ell+1)/r^2$. The equations presented in this section have $\mathcal{E}_{ab}$ in the right-hand-side; one can substitute $\mathcal{E}_{ab}=8\,\pi\,\mathcal{T}_{ab}$ and use the relevant source for the problem concerned (here $\mathcal{T}_{ab}$ is the stress-energy tensor). The coordinate-form of  equation \eqref{new_eqn} is presented in App.\ \ref{app:coord}.

\section{Following Chandrasekhar's Analysis}\label{FCA_1}
Chandrasekhar's work finding relations between solutions to equations is especially interesting in this context because he was able to relate solutions of the even parity equation to solutions of the odd parity equations, even though these sectors are completely decoupled.  He was in fact, relating solution spaces for the equations in question, and not particular solutions for one parity to concurrent solutions for the other parity for a given metric perturbation.

With our odd- and even-parity invariants now presented, along with the hyperbolic equations they solve, we proceed to follow Chandrasekhar and first write the homogenous Regge-Wheeler and Zerilli equations as one, but also in GHP form.  To do so, we will first need to introduce a spin and boost weight scalar $\zeta$ which satisfies $\zeta\propto(-\psi_2)^{-1/3}$, where the constant is chosen so the the Minkowski space limit is well defined (and becomes exactly $r$ in polar coordinates)\cite{AA2}.  Then we introduce new odd parity variables $I_{O_{1a}}/\psi_2$ (\ie the imaginary fractional change in $\psi_2$) and $\zeta^4 I_{O_{2}}$, which will both satisfy equation (\ref{CMEq}) below for $I_{-}$ (Note: whereas these two new quantities will each satisfy the same equation and have the same physical dimension, only one of them remains regular in the flat space limit.), while for even parity, with $I_{+}$ also satisfying equation (\ref{CMEq}), then the quantity $\zeta^{-4}(\hat{\mathcal{X}}-4\rho\rho'+2\psi_2)I_{+}$ will satisfy the Zerilli equation for $I_Z$. 

Following Chandrasekhar's work in \cite{Chandra}, one can write the homogenous equations for $I_\pm$, defined using the transformations above, with Chandrasekhar's master equation being given by
\begin{align}\label{CMEq}
\Big[\zeta\big( \rho\,\kata^\prime + \rho^\prime\kata \big)\Big]^2 I_{\pm} - \Big[\zeta\big( \rho\,\kata^\prime - \rho^\prime\kata \big)\Big]^2 I_{\pm} = V_{\pm} \, I_{\pm},
\end{align}
where the potential $V_\pm$ is given by
\begin{widetext}
\begin{eqnarray}
\zeta^{-2}\big( \hat{\mathcal{X}}-4\,\rho\,\rho^\prime + 2\,\psi_2 \big)^2 \,V_{\pm} &=&\pm24\,\rho\,\rho^\prime\psi_2 \Big[ \psi_2\big( \hat{\mathcal{X}}+2\,\psi_2 \big) -2\,\rho\,\rho^\prime \big( \hat{\mathcal{X}}+\psi_2 \big) + 8\,\rho^2 \rho^{\prime 2}\Big] + 144\,\rho^2\rho^{\prime 2}\psi_2^{2} \nonumber \\
&&+ 2\,\rho\,\rho^\prime \hat{\mathcal{X}}\,\big( \hat{\mathcal{X}}-4\,\rho\,\rho^\prime -4\,\psi_2 \big)\,\big( \hat{\mathcal{X}}-4\,\rho\,\rho^\prime +2\,\psi_2 \big).
\end{eqnarray}
\end{widetext}

Once solutions to the equations for the odd- and even-parity invariants are written in this form, one can relate the solution spaces for the two invariants, $I_+$ and $I_-$, as follows,
\begin{widetext}
\begin{eqnarray} \label{Ipm_relation}
\zeta^{-4} \big( \hat{\mathcal{X}}-4\,\rho\,\rho^\prime + 2\,\psi_2 \big)\, I_{\pm} &=& \Big[ \hat{\mathcal{X}}\,\big( \hat{\mathcal{X}}-4\,\rho\,\rho^\prime -4\,\psi_2 \big)\,\big( \hat{\mathcal{X}}-4\,\rho\,\rho^\prime + 2\,\psi_2 \big) + 144\,\rho\,\rho^\prime\psi_2^2\Big]\,I_{\mp} \nonumber \\
&& \mp 12\,\psi_2 \, \big( \hat{\mathcal{X}}-4\,\rho\,\rho^\prime + 2\,\psi_2 \big) \, \big( \rho\,\kata^\prime + \rho^\prime\kata \big)\, I_\mp\,.
\end{eqnarray}
\end{widetext}

One can also go on and relate these invariants to the perturbed Weyl scalar, $\dot{\psi}_0$,
\begin{align}
\dot{\psi}_0 = \edt\,\edt \, h_{11} + \kata\,\kata\,h_{33} - 2\,\rho\,\kata\,h_{33} - 2\,\edt\,\kata\,h_{13} + 2\,\rho\,\edt\,h_{13}.
\end{align}
in the source-free region. $\dot{\psi}_0$ which can be \emph{scalarized} to
\begin{widetext}
\begin{eqnarray}
\rho^\prime \rho^\prime \edt^\prime \edt^\prime \dot{\psi}_0 &=& \frac14 \hat{\mathcal{X}}\,\big( \hat{\mathcal{X}}-4\,\rho\,\rho^\prime -4\,\psi_2 \big)\,\rho^{\prime 2} h_{11} \nonumber \\ 
&& - \frac12 \big( \hat{\mathcal{X}}-4\,\rho\,\rho^\prime - 4\,\psi_2 \big) \Big[ \big( \rho\,\kata^\prime + \rho^\prime\kata \big) - \big( \rho\,\kata^\prime - \rho^\prime\kata \big) - 6\,\rho\,\rho^\prime +2\,\psi_2  \Big] \, \rho^\prime \edt^\prime h_{13} \nonumber \\
&& + \frac14 \Big[ \big( \rho\,\kata^\prime + \rho^\prime\kata \big)^2 + \big( \rho\,\kata^\prime - \rho^\prime\kata \big)^2 - 2\, \big( \rho\,\kata^\prime + \rho^\prime\kata \big)\, \big( \rho\,\kata^\prime - \rho^\prime\kata \big) + 2\,\big( \psi_2-7\,\rho\,\rho^\prime \big)\,\big( \rho\,\kata^\prime + \rho^\prime\kata \big) \nonumber \\
&& \qquad +\big( 16\,\rho\,\rho^\prime - 2\,\psi_2\big) \big( \rho\,\kata^\prime - \rho^\prime\kata \big) +24\,\rho^2\rho^{\prime 2}\Big] \, \edt^\prime \edt^\prime \, h_{33}.
\end{eqnarray}
\end{widetext}
The homogenous equation it solves is
\begin{align}
&\Big[\zeta\big( \rho\,\kata^\prime + \rho^\prime\kata \big)\Big]^2 Y_0 - \Big[\zeta\big( \rho\,\kata^\prime - \rho^\prime\kata \big)\Big]^2 Y_0 -\, \zeta \big( 4\,\psi_2-8\,\rho\,\rho^\prime \big)\,\Lambda_-\,Y_0 \nonumber \\
&= \zeta^2\Big[ 2\,\rho\,\rho^\prime \big( \hat{\mathcal{X}}-4\,\rho\,\rho^\prime - 4\,\psi_2 \big) + 12\,\rho\,\rho^\prime\psi_2\Big]\,Y_0
\end{align}
where (note, again, that the flat space limit is not regular here)
\begin{align}
\psi_{2}Y_0 &= \zeta^2\rho^\prime\, \rho^\prime\, \edt^\prime \,\edt^\prime \,\dot{\psi}_0, \quad\textrm{and} \\ 
\Lambda_\pm &= \zeta \Big[ \big( \rho\,\kata^\prime + \rho^\prime\kata \big) \mp \big( \rho\,\kata^\prime - \rho^\prime\kata \big) \Big]. 
\end{align}
Once in this form, solutions to the equation for $Y_0$ can be related to solutions for the equations satisfied by $I_\pm$,
\begin{align}
\mathcal{F} \, \mathcal{F} \, Y_0 &= \mathcal{V}_\pm\, I_\pm - \mathcal{F}\left( \mathcal{W}_\pm + 2\,\mathcal{F}\,\mathscr{T} \right) \Lambda_+ I_\pm \label{Y0:Ipm}\\
\mathcal{F}\, \mathcal{K}_\pm (2\,\rho\,\rho^\prime)^2 I_\pm &= (2\,\rho\,\rho^\prime)^2\,\mathcal{F}\,\mathcal{F}\,Y_0 + \left( \mathcal{W}_\pm + 2\,\mathcal{F}\,\mathscr{T} \right) \Lambda_- Y_0,\label{Ipm:Y0}
\end{align}
where
\begin{align}
2\,\rho\,\rho^\prime\,\mathcal{F} &= \zeta^2 \big( \hat{\mathcal{X}}-4\,\rho\,\rho^\prime +2\,\psi_2 \big), \\
\mathcal{K}_\pm &= \kappa \pm 2\,\beta \,\mathscr{T}, \\ 
\kappa &= \zeta^4 \, \hat{\mathcal{X}}\,\big( \hat{\mathcal{X}}-4\,\rho\,\rho^\prime - 4\,\psi_2 \big), \\
\,\rho\,\rho^\prime\,\zeta^{-3}\mathcal{W}_\pm &= \big( 2\,\rho\,\rho^\prime-\psi_2 \big)\,\hat{\mathcal{X}}-2\,\psi_2^2-8\,\rho^2\rho^{\prime 2} +2\rho\,\rho^\prime\psi_2\mp\beta\rho\,\rho^\prime\zeta^{-3}, \\ 
\mathcal{V}_\pm &= \mathcal{F} \, \kappa\pm \beta \Big[ \zeta\big( \rho\,\kata^\prime + \rho^\prime\kata \big)\mathcal{F} \pm \beta \Big], \\ 
\beta &= 6\,\psi_2\,\zeta^3, \\ 
\mathscr{T} &= \zeta \big( \rho\,\kata^\prime - \rho^\prime\kata \big).
\end{align}
Its prime relating \emph{scalarized} solutions for $\dot{\psi}_4$ with solutions for $I_\pm$ also holds. We emphasize again that one should be very careful when using Eqs (\ref{Ipm_relation}, \ref{Y0:Ipm} and \ref{Ipm:Y0}) since these transformations should be used for solution classes of the equations specified, and not for particular parts of the solution corresponding to a given metric perturbation. 
\section{Other gauge-invariants}
\label{sec:recipe}

To systematically get more gauge-invariants, we begin with looking for $2^{nd}$ order invariants other than $\dot{\psi_0}$, $\dot{\psi_4}$, $I_{O_{1a}}$ and $\mathcal{E}_{ab}$'s. It can now be shown that these are the only $2^{nd}$ order invariants possible, and any more of type (0,0) invariants would just be linear combinations of their scalarized version. To find other (higher order) gauge-invariants we translate the gauge transformations given in Eq \eqref{gaugetransf} to the ones for the metric components given by Regge-Wheeler-Zerilli in terms of radial components of tensorial spherical harmonics as follows (but without the separation in terms of radial and angular components),
\begin{align} \label {gaugetransfRWZ}
\rho^{\prime 2}(\delta_\xi h)_{11} - \rho^2(\delta_\xi h)_{22} &= -\left(\rho\,\kata^\prime - \rho^\prime\kata\right)\,\left( \rho^\prime \xi_1 + \rho\, \xi_2\right) \nonumber \\
&\ \ \ +\left(\rho\,\kata^\prime + \rho^\prime\kata-2\,\rho\,\rho^\prime+2\,\psi_2\right)\,\left( \rho^\prime \xi_1 - \rho\, \xi_2\right) \nonumber \\
\rho^{\prime 2}(\delta_\xi h)_{11} - 2\,\rho\,\rho^\prime(\delta_\xi h)_{12}+\rho^2(\delta_\xi h)_{22} &= 2\,\psi_2\left( \rho^\prime \xi_1 + \rho\, \xi_2\right) -\left(\rho\,\kata^\prime - \rho^\prime\kata\right)\,\left( \rho^\prime \xi_1 - \rho\, \xi_2\right) \nonumber \\
\rho^{\prime 2}(\delta_\xi h)_{11} + 2\,\rho\,\rho^\prime(\delta_\xi h)_{12}+\rho^2(\delta_\xi h)_{22} &= \left(\rho\,\kata^\prime + \rho^\prime\kata-2\,\rho\,\rho^\prime +\,\psi_2\right)\,\left( \rho^\prime \xi_1 + \rho\, \xi_2\right) \nonumber \\
\edt^\prime \edt'(\delta_\xi h)_{33}+\edt\,\edt\,(\delta_\xi h)_{44} &= \left( \edt\,\edt' +\edt'\edt-4\,\rho\,\rho'-4\,\psi_2\right)  \left( \edt'\xi_3+\edt\,\xi_4 \right)  \nonumber \\
\rho'\edt'(\delta_\xi h)_{13} +\rho'\edt \,(\delta_\xi h)_{14} + \rho\,\edt'(\delta_\xi h)_{23} +\rho\,\edt\,(\delta_\xi h)_{24} &= \left( \edt\,\edt' +\edt'\edt\right) \left( \rho'\xi_1+\rho\,\xi_2 \right) + \left( \rho\,\kata'+\rho'\kata \right) \left( \edt'\xi_3+\edt\,\xi_4 \right) \nonumber \\
\rho'\edt'(\delta_\xi h)_{13} +\rho'\edt \,(\delta_\xi h)_{14} - \rho\,\edt'(\delta_\xi h)_{23} - \rho\,\edt \,(\delta_\xi h)_{24} &= \left( \edt\,\edt' +\edt'\edt\right) \left( \rho'\xi_1-\rho\,\xi_2 \right) - \left( \rho\,\kata'-\rho'\kata \right) \left( \edt'\xi_3+\edt\,\xi_4 \right) \nonumber \\
(\delta_\xi h)_{34} &= 2 \left( \rho'\xi_1+\rho\,\xi_2 \right) + \left( \edt'\xi_3+\edt\,\xi_4 \right) \nonumber \\
\rho'\edt'(\delta_\xi h)_{13} -\rho'\edt \,(\delta_\xi h)_{14} + \rho\,\edt'(\delta_\xi h)_{23} -\rho\,\edt\,(\delta_\xi h)_{24} &= \left( \rho\,\kata'+\rho'\kata \right) \left( \edt'\xi_3-\edt\,\xi_4 \right) \nonumber \\
\rho'\edt'(\delta_\xi h)_{13} -\rho'\edt \,(\delta_\xi h)_{14} - \rho\,\edt'(\delta_\xi h)_{23} +\rho\,\edt\,(\delta_\xi h)_{24} &= -\left( \rho\,\kata'-\rho'\kata \right) \left( \edt'\xi_3-\edt\,\xi_4 \right) \nonumber \\
\edt^\prime \edt'(\delta_\xi h)_{33}-\edt\,\edt\,(\delta_\xi h)_{44} &= \left( \edt\,\edt' +\edt'\edt-4\,\rho\,\rho'-4\,\psi_2\right) \left( \edt'\xi_3-\edt\,\xi_4 \right). 
\end{align}

\subsection{Odd Parity Invariants}\label{OPI_2}
We begin with finding the odd-parity gauge-invariants from the above equations. 
The last three 
of eqs \eqref{gaugetransfRWZ}, are of odd-parity.  Hence, there are three ways to eliminate the odd-parity gauge vector projection, $\left( \edt'\xi_3-\edt\,\xi_4 \right)$, to obtain the following invariants (here and below, names chosen are consistent with those used in ref.\ \cite{Chen:2016plo}),
\begin{widetext}
\begin{align}
I_\beta=&\left( \rho\,\kata'-\rho'\kata \right)\left( \edt'\edt'h_{33} - \edt\,\edt\,h_{44} \right) + \left( \edt\,\edt' +\edt'\edt-4\,\rho\,\rho'-4\,\psi_2\right) \left( \rho'\edt' h_{13} -\rho'\edt \,h_{14} - \rho\,\edt' h_{23} +\rho\,\edt\, h_{24} \right), \nonumber \\
I_\alpha=&\left( \rho\,\kata'+\rho'\kata-4\,\rho\,\rho' \right)\left( \edt'\edt'h_{33} - \edt\,\edt\,h_{44} \right) - \left( \edt\,\edt' +\edt'\edt-4\,\rho\,\rho'-4\,\psi_2\right) \left( \rho'\edt' h_{13} -\rho'\edt \,h_{14} + \rho\,\edt' h_{23} -\rho\,\edt\, h_{24} \right), \nonumber \\
I_\gamma=&\left( \rho\,\kata'-\rho'\kata\right)\left( \rho'\edt' h_{13} -\rho'\edt \,h_{14} + \rho\,\edt' h_{23} -\rho\,\edt\, h_{24} \right) + \left( \rho\,\kata'+\rho'\kata-2\,\rho\,\rho' \right) \left( \rho'\edt' h_{13} -\rho'\edt \,h_{14} - \rho\,\edt' h_{23} +\rho\,\edt\, h_{24} \right).
\end{align}
\end{widetext}
It will be recognized that $I_\gamma\equiv I_{O1}$ given by eq.\ \eqref{IO1} and $I_\alpha\equiv-I_{O2}$ given by eq.\ \eqref{IO2}, which have already been related to $\Im{\dot{\psi}_2}$ and its time derivative in section \ref{OPI_1}.  It will be seen below that $I_\beta$, the first in the list above, is related to the radial derivative of $\Im{\dot{\psi}_2}$, modulo Einstein tensor components and a multiplicative factor. The exact relation is presented in section \ref{CInv} below.  One should also note that $I_\alpha$ and $I_\gamma$ each obey a second-order hyperbolic PDE, while $I_\beta$ (proportional to the radial derivative of $I_\gamma$) does not.  However, three invariants are not all independent:
\begin{widetext}
\begin{align}
&\left(\rho\,\kata^\prime+\rho^\prime\kata -6\,\rho\,\rho^\prime \right)I_\beta - \left(\rho\,\kata^\prime-\rho^\prime\kata \right)I_\alpha - \left( \edt\,\edt' +\edt'\edt-4\,\rho\,\rho'-4\,\psi_2\right)I_\gamma = 0, \\
\end{align}
\end{widetext}
which is exactly equivalent to the Regge-Wheeler equation \eqref{RW}.

\subsection{Even Parity Invariants}\label{EPI_2}
To get the even-parity invariants, we concentrate on the first seven of Eqs \eqref{gaugetransfRWZ}. Instead of just one, we now have three even-parity gauge vectors projections, $\left( \rho'\xi_1+\rho\,\xi_2 \right)$, $\left( \rho'\xi_1-\rho\,\xi_2 \right)$ and $\left( \edt'\xi_3+\edt\,\xi_4 \right)$. The recipe involves eliminating these three quantities step-by-step. For example, one can use the fifth and seventh equations of the set and eliminate $\left( \edt'\xi_3+\edt\,\xi_4 \right)$, and finally use the third to eliminate $\left( \rho'\xi_1+\rho\,\xi_2 \right)$. Obviously, a number of combinations is possible, potentially leading to a variety of invariants that can be calculated from these seven equations. We choose to translate Steven Detweiler's systematic coordinate-dependent work into a coordinate-independent, covariant GHP formulation. The first step is to write three gauge-vectors in terms of metric perturbations with as few derivatives as possible, and we choose the fourth, sixth and seventh of Eqs \eqref{gaugetransfRWZ},
\begin{widetext}
\begin{align}
\left( \edt\,\edt' +\edt'\edt-4\,\rho\,\rho'-4\,\psi_2\right)\left( \edt\,\edt' +\edt'\edt\right)\left( \edt^\prime\xi_3 + \edt\,\xi_4 \right) &=\left( \edt\,\edt' +\edt'\edt\right)\left( \edt^\prime \edt' (\delta_\xi\, h)_{33}+\edt\,\edt\,(\delta_\xi\, h)_{44}\right), \\
\left( \edt\,\edt' +\edt'\edt-4\,\rho\,\rho'-4\,\psi_2\right)\left( \edt\,\edt' +\edt'\edt\right)\left( \rho^\prime\xi_1 - \rho\,\xi_2 \right) & =\left( \rho\,\kata'-\rho'\kata \right) \left( \edt^\prime \edt' (\delta_\xi\, h)_{33}+\edt\,\edt\,(\delta_\xi\, h)_{44}\right)\nonumber \\
+\left( \edt\,\edt' +\edt'\edt-4\,\rho\,\rho'-4\,\psi_2\right)& \left( \rho'\edt' (\delta_\xi\, h)_{13} +\rho'\edt \,(\delta_\xi\, h)_{14} - \rho\,\edt' (\delta_\xi\, h)_{23} - \rho\,\edt \,(\delta_\xi\, h)_{24} \right), \\
\left( \edt\,\edt' +\edt'\edt-4\,\rho\,\rho'-4\,\psi_2\right)\left( \edt\,\edt' +\edt'\edt\right)\left( \rho^\prime\xi_1 + \rho\,\xi_2 \right) &= \frac12\left( \edt\,\edt' +\edt'\edt-4\,\rho\,\rho'-4\,\psi_2\right)\left( \edt\,\edt' +\edt'\edt\right) (\delta_\xi\, h)_{34} \nonumber \\
& - \frac12 \left( \edt\,\edt' +\edt'\edt\right)\left( \edt^\prime \edt' (\delta_\xi\, h)_{33}+\edt\,\edt\,(\delta_\xi\, h)_{44}\right).
\end{align}
\end{widetext}
We use these to eliminate the gauge vectors projections from the first, second, third and fifth of Eqs \eqref{gaugetransfRWZ}, to have 4 even-parity invariants, 
\begin{widetext}
\begin{align}
I_\delta =\,&2 \left( \edt\,\edt' +\edt'\edt-4\,\rho\,\rho'-4\,\psi_2\right)\left( \edt\,\edt' +\edt'\edt\right)\left( \rho^{\prime 2}h_{11} - \rho^2h_{22} \right) \nonumber \\
&+ \left( \edt\,\edt' +\edt'\edt-4\,\rho\,\rho'-4\,\psi_2\right)\left( \edt\,\edt' +\edt'\edt\right)\left( \rho\,\kata'-\rho'\kata \right)h_{34} \nonumber \\
&-2 \left( \edt\,\edt' +\edt'\edt-4\,\rho\,\rho'-4\,\psi_2\right)\left( \rho\,\kata'+\rho'\kata -6\,\rho\,\rho^\prime+2\,\psi_2\right) \left( \rho'\edt' h_{13} +\rho'\edt \,h_{14} - \rho\,\edt' h_{23} - \rho\,\edt \,h_{24} \right) \nonumber \\
&-\left( \edt\,\edt' +\edt'\edt-4\,\rho\,\rho'-4\,\psi_2\right) \left( \rho\,\kata'-\rho'\kata \right)\left( \edt^\prime \edt' h_{33}+\edt\,\edt\,h_{44}\right) \nonumber \\
&-2\left( \rho\,\kata'-\rho'\kata \right)\left( \rho\,\kata'+\rho'\kata-6\,\rho\,\rho^\prime+4\,\psi_2 \right) \left( \edt^\prime \edt' h_{33}+\edt\,\edt\,h_{44}\right), \\
I_\epsilon =\,& \left( \edt\,\edt' +\edt'\edt-4\,\rho\,\rho'-4\,\psi_2\right)\left( \edt\,\edt' +\edt'\edt\right)\left( \rho^{\prime 2}h_{11} -2\,\rho\,\rho^\prime h_{12}+ \rho^2h_{22} \right) \nonumber \\
&-\psi_2 \left( \edt\,\edt' +\edt'\edt-4\,\rho\,\rho'-4\,\psi_2\right)\left( \edt\,\edt' +\edt'\edt\right)h_{34} \nonumber \\
&+2 \left( \edt\,\edt' +\edt'\edt-4\,\rho\,\rho'-4\,\psi_2\right)\left( \rho\,\kata'-\rho'\kata\right) \left( \rho'\edt' h_{13} +\rho'\edt \,h_{14} - \rho\,\edt' h_{23} - \rho\,\edt \,h_{24} \right) \nonumber \\
&+\psi_2\left( \edt\,\edt' +\edt'\edt\right) \left( \edt^\prime \edt' h_{33}+\edt\,\edt\,h_{44}\right) \nonumber \\
&+2\left( \rho\,\kata'-\rho'\kata \right)\left( \rho\,\kata'-\rho'\kata \right) \left( \edt^\prime \edt' h_{33}+\edt\,\edt\,h_{44}\right),\\
I_\chi =\,&- \left( \edt\,\edt' +\edt'\edt-4\,\rho\,\rho'-4\,\psi_2\right)\left( \edt\,\edt' +\edt'\edt\right)h_{34} \nonumber \\
&+2 \left( \edt\,\edt' +\edt'\edt-4\,\rho\,\rho'-4\,\psi_2\right)\left( \rho'\edt' h_{13} +\rho'\edt \,h_{14} - \rho\,\edt' h_{23} - \rho\,\edt \,h_{24} \right) \nonumber \\
&+\left( \edt\,\edt' +\edt'\edt-4\,\rho\,\rho'-4\,\psi_2\right) \left( \edt^\prime \edt' h_{33}+\edt\,\edt\,h_{44}\right) \nonumber \\
&-2\left( \rho\,\kata'+\rho'\kata-6\,\rho\rho^\prime-2\,\psi_2 \right)\left( \edt^\prime \edt' h_{33}+\edt\,\edt\,h_{44}\right), \\
I_\psi=\,& \left( \edt\,\edt' +\edt'\edt-4\,\rho\,\rho'-4\,\psi_2\right)\left( \rho^{\prime 2}h_{11} +2\,\rho\,\rho^\prime h_{12}+ \rho^2h_{22} \right) \nonumber \\
&- \left( \edt\,\edt' +\edt'\edt-4\,\rho\,\rho'-4\,\psi_2\right)\left( \rho\,\kata'+\rho'\kata-2\,\rho\,\rho^\prime+\psi_2 \right)h_{34} \nonumber \\
&+\left( \rho\,\kata'+\rho'\kata-6\,\rho\,\rho^\prime+\psi_2 \right) \left( \edt^\prime \edt' h_{33}+\edt\,\edt\,h_{44}\right).
\end{align}
\end{widetext}
On the other hand, one could have used other components of metric perturbations to eliminate the gauge vector projections, and arrive at a different set of invariants. 

\subsection{Curvature Invariants}\label{CInv}
We now try to relate these odd- and even-parity invariants with curvature invariants. 

\subsubsection{Odd Parity}
We begin with defining three, scalarized, odd-parity invariants 
\begin{align}
{I}_0 &= \rho^\prime \rho^\prime \edt^\prime\edt^\prime\dot{\psi}_0 - \rho^\prime \rho^\prime \edt\,\edt\,\dot{\bar{\psi}}_0,\\
{I}_2 &= \frac{1}{2}(\dot{\psi}_2-\dot{\bar{\psi}}_2)\equiv \Im{\dot{\psi}_2}, \quad \textrm{and}\\
{I}_4 &= \rho\,\rho\,\edt\,\edt\,\dot{\psi}_4 - \rho\,\rho\,\edt^\prime\edt^\prime\dot{\bar{\psi}}_4,
\end{align}
and then seek ways of relating the Detweiler invariants to them.  By direct computation, we find

\begin{widetext}
\begin{align}
%
%
%
%
& I_\alpha = -\left(\rho\,\kata^\prime-\rho^\prime\kata \right)I_2 - 2 \left( \rho'\edt' \mathcal{E}_{13} -\rho'\edt \,\mathcal{E}_{14} + \rho\,\edt' \mathcal{E}_{23} -\rho\,\edt\, \mathcal{E}_{24} \right),\label{Ia} \\
&I_\beta = -\left(\rho\,\kata^\prime+\rho^\prime\kata -8\,\rho\,\rho^\prime \right)I_2 + 2 \left( \rho'\edt' \mathcal{E}_{13} -\rho'\edt \,\mathcal{E}_{14} - \rho\,\edt' \mathcal{E}_{23} + \rho\,\edt\, \mathcal{E}_{24} \right),\label{Ib}\\
& I_\gamma = -2\rho\rho' I_2.\label{Ig}
\end{align}
\end{widetext}
Eqs \eqref{Ia} and \eqref{Ig} together are equivalent to eq.\ \eqref{CPM-RW} above.  Finally, we show that the three curvature invariants are themselves related:
\begin{widetext}
\begin{align}
2\,{I}_0+2\,{I}_4&-\left(\rho\,\kata^\prime-\rho^\prime\kata \right)\left(\rho\,\kata^\prime-\rho^\prime\kata \right)I_2 - \left(\rho\,\kata^\prime+\rho^\prime\kata -10\,\rho\,\rho^\prime+2\,\psi_2 \right)\left(\rho\,\kata^\prime+\rho^\prime\kata -8\,\rho\,\rho^\prime \right)I_2\nonumber\\
& = 2\left(\rho\,\kata^\prime-\rho^\prime\kata \right)\left( \rho'\edt' \mathcal{E}_{13} -\rho'\edt \,\mathcal{E}_{14} + \rho\,\edt' \mathcal{E}_{23} -\rho\,\edt\, \mathcal{E}_{24} \right)\nonumber\\
& - 2\left(\rho\,\kata^\prime+\rho^\prime\kata -10\,\rho\,\rho^\prime+2\,\psi_2 \right)\left( \rho'\edt' \mathcal{E}_{13} -\rho'\edt \,\mathcal{E}_{14} - \rho\,\edt' \mathcal{E}_{23} + \rho\,\edt\, \mathcal{E}_{24}\right). \label{RWEq}
\end{align}
\end{widetext}
To our knowledge, this equation, differentially relating the curvature invariants, $\Im{\dot{\psi}_0}$, $\Im{\dot{\psi}_2}$ and $\Im{\dot{\psi}_4}$, has not previously been explicitly given.  With the analysis on odd-parity invariants complete, we now move to the complicated even-parity invariants. 

\subsubsection{Even Parity}
The even-parity invariants, $I_\delta$, $I_\epsilon$, $I_\chi$ and $I_\psi$, can be related to each other using the following relations,
\begin{widetext}
\begin{align}
4\,{R}_0-4\,{R}_4 &= I_\delta + (\rho\,\kata^\prime - \rho^\prime\kata)I_\chi, \\
2\,{R}_0+ 2\,{R}_4+4\,\rho\,\rho^\prime\left(\edt^\prime\edt^\prime\mathcal{E}_{33}+\edt\,\edt\,\mathcal{E}_{44}\right) &= I_\epsilon-\psi_2I_\chi, \\
2\,{R}_0+ 2\,{R}_4-4\,\rho\,\rho^\prime\left(\edt^\prime\edt^\prime\mathcal{E}_{33}+\edt\,\edt\,\mathcal{E}_{44}\right) &= \left( \edt\,\edt^\prime + \edt^\prime\edt \right)I_\psi-\left( \rho\,\kata^\prime + \rho^\prime\kata - 10\,\rho\,\rho^\prime+\psi_2 \right)I_\chi, \\
4\left( \edt\,\edt^\prime + \edt^\prime\edt - 4\,\rho\,\rho^\prime-4\,\psi_2\right)\left( \rho^{\prime 2}\mathcal{E}_{11} - \rho^2\mathcal{E}_{22} \right) &= I_\delta + \left(\rho\,\kata^\prime - \rho^\prime\kata \right)I_\psi+(\rho\,\kata^\prime - \rho^\prime\kata)XI_\chi, \\
12\,\Psi_2\left( \rho^{\prime 2}\mathcal{E}_{11} - 2\,\rho\,\rho^\prime\mathcal{E}_{12} + \rho^2 \mathcal{E}_{22} \right) + 8\,\rho\,\rho^\prime\Psi_{2} \left( 2\,\mathcal{E}_{12} + \mathcal{E}_{34} \right) &= \left( \edt\,\edt^\prime + \edt^\prime\edt-4\,\rho\,\rho^\prime + 2\,\Psi_2 \right)I_\psi \nonumber \\
&-\left( \rho\,\kata^\prime + \rho^\prime \kata-12\,\rho\,\rho^\prime+\Psi_2 \right)\left( I_\chi+2\,I_\psi \right),
\end{align}
\end{widetext}
where
\begin{align}
{R}_0 &= \rho^\prime \rho^\prime \edt^\prime\edt^\prime\dot{\psi}_0 + \rho^\prime \rho^\prime \edt\,\edt\,\dot{\bar{\psi}}_0, \quad \textrm{and}\\
{R}_4 &= \rho\,\rho\,\edt\,\edt\,\dot{\psi}_4 + \rho\,\rho\,\edt^\prime\edt^\prime\dot{\bar{\psi}}_4 \,\,.
\end{align}
These equations can then be inverted to write the four invariants in terms of curvature invariants as follows
\begin{align}
\left(\rho \,\kata^\prime - \rho^\prime \kata\right) I_\psi &= -4\left(R_0 - R_4\right) + 4\left( \edt\,\edt^\prime + \edt^\prime\edt - 4\,\rho\,\rho^\prime-4\,\psi_2\right)\left(\rho^{\prime 2} \mathcal{E}_{11} - \rho^2 \mathcal{E}_{22}\right), \\
\rho\,\rho^\prime\left(\rho \kata^\prime - \rho^\prime \kata\right)I_\chi &= 6\,\psi_2\left(\rho \, \kata^\prime - \rho^\prime \kata\right)\left(\rho^{\prime 2} \mathcal{E}_{11} -2\,\rho\rho^\prime\mathcal{E}_{12} + \rho^2 \mathcal{E}_{22}\right) + 4\,\rho\,\rho^\prime\psi_2 \left(\rho \,\kata^\prime - \rho^\prime \kata\right)\left( 2 \,\mathcal{E}_{12}+\mathcal{E}_{34} \right) \nonumber \\
&- \left(\rho \,\kata^\prime - \rho^\prime \kata\right)\left(R_0 + R_4\right) + 2\,\rho\,\rho^\prime\left(\rho \,\kata^\prime - \rho^\prime \kata\right)\left( \edt^\prime \edt^\prime \mathcal{E}_{33} + \edt\,\edt\,\mathcal{E}_{44} \right) \nonumber \\
&- 4\left(\rho \,\kata^\prime + \rho^\prime \kata-12\,\rho\,\rho^\prime\right) \left( R_0 - R_4 \right) \nonumber \\
&+ 4 \left(\rho\, \kata^\prime + \rho^\prime \kata-12\,\rho\,\rho^\prime\right)\left( \edt\,\edt^\prime + \edt^\prime\edt - 4\,\rho\,\rho^\prime-4\,\psi_2\right)\left(\rho^{\prime 2} \mathcal{E}_{11} - \rho^2 \mathcal{E}_{22}\right), \\
\rho\,\rho^\prime I_\delta &= 4\,\rho\,\rho^\prime\left(R_0 - R_4\right) - 6\,\psi_2\left(\rho \,\kata^\prime - \rho^\prime \kata\right)\left(\rho^{\prime 2} \mathcal{E}_{11} -2\,\rho\,\rho^\prime\mathcal{E}_{12} + \rho^2 \mathcal{E}_{22}\right) \nonumber \\
&- 4\,\rho\,\rho^\prime\psi_2 \left(\rho \,\kata^\prime - \rho^\prime \kata\right)\left( 2 \,\mathcal{E}_{12}+\mathcal{E}_{34} \right) +  \left(\rho \,\kata^\prime - \rho^\prime \kata\right) \left( R_0 + R_4 \right) \nonumber \\
&-2\,\rho\,\rho^\prime\left(\rho \,\kata^\prime - \rho^\prime \kata\right)\left( \edt^\prime \edt^\prime \mathcal{E}_{33} + \edt\,\edt\,\mathcal{E}_{44} \right)+ 4\left(\rho \,\kata^\prime + \rho^\prime \kata-12\,\rho\,\rho^\prime\right) \left( R_0 - R_4 \right) \nonumber \\
&- 4 \left(\rho\, \kata^\prime + \rho^\prime \kata-12\,\rho\,\rho^\prime\right)\left( \edt\,\edt^\prime + \edt^\prime\edt - 4\,\rho\,\rho^\prime-4\,\psi_2\right)\left(\rho^{\prime 2} \mathcal{E}_{11} - \rho^2 \mathcal{E}_{22}\right), \\
\rho\,\rho^\prime\left(\rho \,\kata^\prime - \rho^\prime \kata\right)I_\epsilon &= \left(2\,\rho\,\rho^\prime-\psi_2\right)\left(\rho\, \kata^\prime - \rho^\prime \kata\right) \left( R_0 + R_4\right)+4\,\rho^2\rho^{\prime 2}\left(\rho \,\kata^\prime - \rho^\prime \kata\right)\left( \edt^\prime \edt^\prime \mathcal{E}_{33} + \edt\,\edt\,\mathcal{E}_{44} \right) \nonumber \\
&+ 6\,\psi_2^2\left(\rho\, \kata^\prime - \rho^\prime \kata\right)\left(\rho^{\prime 2} \mathcal{E}_{11} -2\,\rho\,\rho^\prime\mathcal{E}_{12} + \rho^2 \mathcal{E}_{22}\right) + 4\,\rho\,\rho^\prime\psi_2^2\left(\rho \,\kata^\prime - \rho^\prime \kata\right)\left( 2 \,\mathcal{E}_{12}+\mathcal{E}_{34} \right) \nonumber \\
&+ 2\,\rho\,\rho^\prime\psi_2\left(\rho \,\kata^\prime - \rho^\prime \kata\right)\left( \edt^\prime \edt^\prime \mathcal{E}_{33} + \edt\,\edt\,\mathcal{E}_{44} \right) - 4\,\psi_2 \left(\rho \,\kata^\prime + \rho^\prime \kata-12\,\rho\,\rho^\prime\right) \left( R_0 - R_4 \right) \nonumber \\
&+ 4\psi_2 \left(\rho \,\kata^\prime + \rho^\prime \kata-12\,\rho\,\rho^\prime\right)\left( \edt\,\edt^\prime + \edt^\prime\edt - 4\,\rho\,\rho^\prime-4\,\psi_2\right)\left(\rho^{\prime 2} \mathcal{E}_{11} - \rho^2 \mathcal{E}_{22}\right).
\end{align}
Along with these we also get an equation relating the curvature invariants with Einstein tensors,
\begin{align}
&-\left(\rho \,\kata^\prime - \rho^\prime \kata\right)\left(\rho \,\kata^\prime + \rho^\prime \kata-16\,\rho\,\rho^\prime+3\,\psi_2\right) \left( R_0+R_4\right) +4\,\rho\,\rho^\prime \left( \edt\,\edt^\prime + \edt^\prime\edt \right)  \left( R_0-R_4\right) \nonumber \\
&- 4 \left(\rho \,\kata^\prime + \rho^\prime \kata-16\,\rho\,\rho^\prime+3\,\psi_2\right) \left(\rho \,\kata^\prime + \rho^\prime \kata-12\,\rho\,\rho^\prime\right)  \left( R_0-R_4\right) \nonumber \\
&=4\,\rho^2\rho^{\prime 2} \left(\rho \,\kata^\prime - \rho^\prime \kata\right) \left( \edt^\prime \edt^\prime \mathcal{E}_{33} + \edt\,\edt\,\mathcal{E}_{44} \right) + 4\,\rho\,\rho^\prime  \left( \edt\,\edt^\prime + \edt^\prime\edt \right)\left( \edt\,\edt^\prime + \edt^\prime\edt - 4\,\rho\,\rho^\prime-4\,\psi_2\right)\left(\rho^{\prime 2} \mathcal{E}_{11} - \rho^2 \mathcal{E}_{22}\right) \nonumber \\
&+\left(\rho \,\kata^\prime + \rho^\prime \kata-16\,\rho\,\rho^\prime+3\,\psi_2\right) \Big[  -6\,\psi_2 \left(\rho \,\kata^\prime - \rho^\prime \kata\right)\left(\rho^{\prime 2} \mathcal{E}_{11} -2\,\rho\,\rho^\prime\mathcal{E}_{12} + \rho^2 \mathcal{E}_{22}\right) - 4\,\rho\,\rho^\prime\psi_2\left(\rho \,\kata^\prime - \rho^\prime \kata\right) \left( 2 \,\mathcal{E}_{12}+\mathcal{E}_{34} \right) \nonumber \\
&-2\,\rho\,\rho^\prime \left(\rho \,\kata^\prime - \rho^\prime \kata\right)\left( \edt^\prime \edt^\prime \mathcal{E}_{33} + \edt\,\edt\,\mathcal{E}_{44} \right) - 4 \left(\rho \,\kata^\prime + \rho^\prime \kata-12\,\rho\,\rho^\prime\right)\left( \edt\,\edt^\prime + \edt^\prime\edt - 4\,\rho\,\rho^\prime-4\,\psi_2\right)\left(\rho^{\prime 2} \mathcal{E}_{11} - \rho^2 \mathcal{E}_{22}\right) \Big].
\end{align}
Its nature is yet to be investigated.
Though none of these 4 even-parity invariants solve a $2^{nd}$-order hyperbolic PDE, their combinations, the Zerilli invariant, ${I}_{Z}$ and the new even-parity invariant, ${I}_{\nu E1}$, which are related to these invariants as follows,
\begin{align}
2\,{I}_{\nu E1} &= 2I_\psi + I_\chi, \\
2\,{I}_{Z} &= I_\delta - \left( \rho\,\kata^\prime - \rho^\prime \kata \right) I_\chi,
\end{align}
do each obey a second order hyperbolic PDE as shown in Eqs (\eqref{new_eqn},  \eqref{eq:zereq}) above. 
\section{Conclusions and outlook}
\label{sec:concl}

This work, which started as a \emph{warm-up exercise} to study perturbations on Kerr spacetime, has led to many new and interesting results. 
In this article, we investigate gauge-invariant quantities in Schwarzschild spacetime and, where possible, present and derive a $2^{nd}$-order hyperbolic PDE they obey. Previous work by numerous authors who studied perturbations on Schwarzschild background have been translated to a covariant, coordinate-independent and tetrad-rotation invariant form using the GHP-tools. Relations between the various odd- and even-parity invariants, and curvature invariants that enter these relations, have been derived and presented here as a set of minimal number of equations required. While working on this, we discovered a new even-parity invariant which obeys a second-order hyperbolic PDE.  Both the invariant, and the equation it obeys, are of order lesser than Zerilli's. 
It is interesting to note how the real and imaginary parts of the Weyl scalars are related to the various invariants derived in this work. Chandra's work on relating the RWZ-invariants with each other and the Weyl scalars have also been translated to a GHP-form, and one should be careful when using these relations as these relations relate elements of the solution set and not solutions satisfying the same boundary conditions. In the Appendices, we show how to translate the metric perturbation in GHP form to that in the RWZ-gauge. We also present the Einstein tensors and their $scalarized$ version which can easily be related to ones in the works of RWZ and others. 

A number of difficulties come up when carrying forward this work to studying perturbations of Kerr black hole, and deriving gauge-invariant quantities and the equations they obey. The obvious one is the presence of complex Ricci coefficients, $\pi$ and $\tau$ (along with $\rho)$; and $\psi_2$ being complex. Commutation relations between various operators are not simple anymore, for example, $\edt$ and $\edt^\prime$ (when acting on a quantity of type (0,0)) don't commute with each other, nor can $\edt$ or $\edt^\prime$ pass through $\rho$ or $\psi_2$ anymore. Angular operators like $(\edt\,\edt^\prime + \edt^\prime\edt)$ that gives a term proportional to $\ell(\ell+1)$, and $(\edt\,\edt^\prime + \edt^\prime\edt - 4\,\rho\,\rho^\prime - 4\,\psi_2)$ that is proportional to $(\ell-1)(\ell+2)$, are no more possible in Kerr spacetime; if needed, one will have to use the Teukolsky-Starobinsky identities. With these difficulties comes a lot of freedom, for example, there is more than one way to write an operator proportional to $\partial_r$ or $\partial_t$. Unlike in this work, many combinations are possible to scalarize the metric perturbations, Weyl scalars and Einstein tensors, for example, one will have to use a linear combination of $\edt$, $\tau$ and $\bar{\tau}^\prime$ to get a quantity of type (1,-1), where as in this work, the only possibility was the $\edt$-operator (owing to $\tau=0$ here) which is also a spin-raising operator for spin-weighted spherical harmonics. While struggling with these difficulties and exploring the various possibilities in defining quantities of interest (owing to the freedom involved in Kerr spacetime), we hope to derive quantities equivalent to the odd- and even-parity invariants in Schwarzschild spacetime and the equations they obey. 

\acknowledgments

This work was supported in part by the European Research Council under the European Union's Seventh Framework Programme (FP7/2007-2013)/ERC grant agreement no. 304978. AGS would like to thank IHES for hospitality at various stages of development of this work.  This work was supported by NSF Grants PHY 1205906 and PHY 1314529 to UF.  BFW acknowledges visitor support from the Albert Einstein Institute, Golm, the University of Southampton, the Institut d'Astrophysique de Paris, the Universit\'e Paris Diderot, and CNRS via the IAP and APC, through each of which part of this work was carried out.  We acknowledge the financial support from the UnivEarthS Labex program at Sorbonne Paris Cit\'e (ANR-10-LABX-0023 and ANR-11-IDEX-0005-02) and from the Agence Nationale de la Recherche through the grant ANR-14-CE03-0014-01. LA was supported in part by the Wallenberg foundation, through a grant to the Royal Institute of Technology (KTH, Stockholm). We are grateful to Royal Institute of Technology (KTH, Stockholm) for hospitality during part of the work on this paper.

\appendix
\section{GHP vs coordinate metric perturbation}
\label{metric_comp}
If one were to use the tensor harmonics given in Appendix A of \cite{Zerilli_PRD}, and the metric perturbation given in Appendix D of \cite{Zerilli_PRD}, we then have the following relations,
\begin{align}
\allowdisplaybreaks
H_{0_{L,M}} &= \frac{r^2}{\left(1-\frac{2M}{r}\right)} \int (\rho^{\prime^2} h_{11} - 2 \rho\rho^\prime h_{12}+\rho^2 h_{22})\bar{Y}_{L,M}d\Omega,  \\ 
H_{2_{L,M}} &= \frac{r^2}{\left(1-\frac{2M}{r}\right)} \int (\rho^{\prime^2} h_{11} + 2 \rho\rho^\prime h_{12}+\rho^2 h_{22})\bar{Y}_{L,M}d\Omega,  \\
H_{1_{L,M}} &= \frac{r^2}{\left(1-\frac{2M}{r}\right)} \int (\rho^{\prime^2} h_{11} - \rho^2 h_{22})\bar{Y}_{L,M}d\Omega,  \\ 
h^{(m)}_{1_{L,M}} &= \frac{-\,r^3}{L(L+1)\left(1-\frac{2M}{r}\right)} \int (\rho\,\edt \,h_{24} + \rho\,\edt^\prime h_{23} + \rho^\prime\edt \,h_{14} + \rho^\prime\edt^\prime h_{13})\bar{Y}_{L,M}d\Omega,  \\
h^{(m)}_{0_{L,M}} &= \frac{\,r^3}{L(L+1)} \int (\rho\,\edt \,h_{24} + \rho\,\edt^\prime h_{23} - \rho^\prime\edt \,h_{14} - \rho^\prime\edt^\prime h_{13})\bar{Y}_{L,M}d\Omega,  \\
G_{L,M} &= \frac{2\,r^2}{(L-1)L(L+1)(L+2)}  \int (\edt^\prime \edt^\prime h_{33} + \edt\,\edt\, h_{44})\bar{Y}_{L,M}d\Omega,  \\
K_{L,M} - \frac{L(L+1)}{2}G_{L,M} &= \int h_{34} \bar{Y}_{L,M}d\Omega,  \\
h_{0_{L,M}} &= \frac{i\,r^3}{L(L+1)}  \int (\rho^\prime\edt^\prime h_{13} - \rho^\prime\edt \,h_{14} - \rho\,\edt^\prime h_{23} + \rho\,\edt \,h_{24})\bar{Y}_{L,M}d\Omega,  \\
h_{1_{L,M}} &= \frac{i\,r^3}{\left(1-\frac{2M}{r}\right)\,L(L+1)} \int (\rho^\prime\edt^\prime h_{13} - \rho^\prime\edt \,h_{14} + \rho\,\edt^\prime h_{23} - \rho\,\edt \,h_{24})\bar{Y}_{L,M}d\Omega,  \\
h_{2_{L,M}} &= \frac{2\,r^4}{(L-1)L(L+1)(L+2)}  \int (\edt^\prime \edt^\prime h_{33} - \edt\,\edt\, h_{44})\bar{Y}_{L,M}d\Omega.
\end{align}
Similar relations hold for Einstein or stress-energy tensors given in Appendix A of \cite{Zerilli_PRD}.
\section{EFE and scalarized versions}
\label{app:EFE}
In this section we present the scalarized versions of the Einstein tensor in terms of metric perturbations. 
\begin{widetext} \allowdisplaybreaks
\begin{align}
\rho^{\prime 2}\mathcal{E}_{11} =& \left( \tderivative \right) \left( \rho^{\prime 2}h_{11} \right) + \frac12 \left( \wacky \right) \left( \rho^{\prime 2}h_{11} \right) \nonumber \\
&- \frac12 \left[ \left( \rderivative \right) - \left( \tderivative \right) - 4\,\rho\,\rho^\prime+2\,\psi_2\right]\left( 2\,\rho\,\rho^\prime h_{12} \right)  \nonumber \\
&+\frac14 \left[ \left( \rderivative \right)^2 + \left( \tderivative \right)^2 - 2\left( \rderivative \right)\left( \tderivative \right)\right] h_{34} \nonumber \\
&+\frac12 \left[ \left( 4\,\rho\,\rho^\prime-\psi_2 \right)\left( \tderivative \right) -\left( 3\,\rho\,\rho^\prime-\psi_2 \right)\left( \rderivative \right)  \right] h_{34} \nonumber \\
&-\frac12 \left[ \left( \rderivative \right) - \left( \tderivative \right) - 6\,\rho\,\rho^\prime+2\,\psi_2\right]\left( \rho^\prime\edt^\prime h_{13} + \rho^\prime\edt\, h_{14} \right), \\
\edt^\prime\edt^\prime \mathcal{E}_{33} =& -\frac18 \left( \bigwacky \right)\left( \wacky \right) \left( 2\,\rho\,\rho^\prime h_{12} \right) \nonumber \\
&+\frac14 \left[ \left( \rderivative \right) + \left( \tderivative \right)-10\,\rho\,\rho^\prime \right] \left( \bigwacky \right)\left( \rho^\prime\edt^\prime h_{13} + \rho\,\edt^\prime h_{23} \right) \nonumber \\
&+\frac14 \left[ \left( \tderivative \right)^2 - \left( \rderivative \right)^2 + 14\,\rho\,\rho^\prime \left(\rderivative \right) -4\,\rho\,\rho^\prime \left( 3\,\rho\,\rho^\prime + \psi_2 \right) \right] \left( \edt^\prime\edt^\prime h_{33} \right), \\
\rho\,\rho^\prime\rho^\prime\edt^\prime \mathcal{E}_{13} =& \frac18 \left( \wacky \right) \left[ \left( \rderivative \right) + \left( \tderivative \right) -4\,\rho\,\rho^\prime \right] \left(  \rho^{\prime 2} h_{11} \right) \nonumber \\
&- \frac{1}{16} \left( \wacky \right) \left[ \left( \rderivative \right) + \left( \tderivative \right) +2\,\psi_2 \right] \left( 2\,\rho\,\rho^\prime h_{12} \right) \nonumber \\
&+ \frac18\rho\,\rho^\prime\left( \wacky \right) \left[ \left( \rderivative \right) - \left( \tderivative \right) \right] h_{34} - \frac14\rho\,\rho^\prime \left( \wacky \right) \left( \rho^\prime\edt\,h_{14} \right) \nonumber \\
&-\frac14\rho\,\rho^\prime \left[ \left( \rderivative \right) - \left( \tderivative \right) - 4\,\rho\,\rho^\prime \right]\left( \edt^\prime\edt^\prime h_{33}\right) \nonumber \\
&+\frac18 \left[ \left( \rderivative \right)^2 + \left( \tderivative \right)^2 - 2\left( \rderivative \right) \left( \tderivative \right) \right] \left( \rho\,\edt^\prime h_{23} \right) \nonumber \\
&+\frac14 \left[ \left( \psi_2 - 7\,\rho\,\rho^\prime \right)\left( \rderivative \right) + \left( 8\,\rho\,\rho^\prime - \psi_2 \right)\left( \rderivative \right) +8\,\rho^2\rho^{\prime 2} \right] \left( \rho\,\edt^\prime h_{23} \right) \nonumber \\
&+\frac18 \left[ \left( \tderivative \right)^2 - \left( \rderivative \right)^2 + 2\,\rho\,\rho^\prime \left( \wacky \right) \right] \left( \rho^\prime\edt^\prime h_{13} \right) \nonumber \\
&+\frac14 \left[ \left( 7\,\rho\,\rho^\prime - \psi_2 \right)\left( \rderivative \right) + \left( 4\,\rho\,\rho^\prime - \psi_2 \right) \left( \tderivative \right) - 8\,\rho\,\rho^\prime \left( 2\,\rho\,\rho^\prime + \psi_2 \right)\right]  \left( \rho^\prime\edt^\prime h_{13} \right), \\
\rho\,\rho^\prime \mathcal{E}_{12} =& \frac12 \left[ \left( \rderivative \right) + \left( \tderivative \right) - 6\,\rho\,\rho^\prime \right] \left( \rho^{\prime 2} h_{11} \right) + \frac12 \left[ \left( \rderivative \right) - \left( \tderivative \right) - 6\,\rho\,\rho^\prime \right] \left( \rho^{2} h_{22} \right) \nonumber \\
&-\frac14  \left(\wacky\right) \left( 2\,\rho\,\rho^\prime h_{12} \right) - \left( \rho\,\rho^\prime + \psi_2 \right)\left( 2\,\rho\,\rho^\prime h_{12} \right) - \frac12 \rho\,\rho^\prime \left( \edt^\prime \edt^\prime h_{33} + \edt\,\edt\,h_{44} \right) \nonumber \\
&+ \frac14 \left[  \left( \rderivative \right) + \left( \tderivative \right) - 10\,\rho\,\rho^\prime \right] \left( \rho^\prime\edt^\prime h_{13} + \rho^\prime \edt\, h_{14} \right) \nonumber \\
&+ \frac14 \left[  \left( \rderivative \right) - \left( \tderivative \right) - 10\,\rho\,\rho^\prime \right] \left( \rho\,\edt^\prime h_{23} + \rho\, \edt\, h_{24} \right) \nonumber \\
&+ \frac14 \left[ \left( \tderivative \right)^2 - \left( \rderivative \right)^2 + 10\,\rho\,\rho^\prime\left( \rderivative \right)  + 2\,\rho\,\rho^\prime \left( \bigwacky \right) \right] h_{34}, \\
\left(\rho\,\rho^\prime\right)^2 \mathcal{E}_{34} =& \frac18\left[ \left( \rderivative \right)^2 + \left( \tderivative \right)^2 + 2\left( \rderivative \right) \left( \tderivative \right)  \right] \left( \rho^{\prime 2} h_{11} \right) \nonumber \\
&-\frac14 \left[ \left( 7\,\rho\,\rho^\prime - \psi_2 \right) \left( \rderivative \right) +  \left( 8\,\rho\,\rho^\prime - \psi_2 \right) \left( \tderivative \right) -12 \left( \rho\,\rho^\prime \right)^2\right] \left( \rho^{\prime 2} h_{11} \right) \nonumber \\
&+\frac18\left[ \left( \rderivative \right)^2 + \left( \tderivative \right)^2 - 2\left( \rderivative \right) \left( \tderivative \right)  \right] \left( \rho^{ 2} h_{22} \right) \nonumber \\
&-\frac14 \left[ \left( 7\,\rho\,\rho^\prime - \psi_2 \right) \left( \rderivative \right) -  \left( 8\,\rho\,\rho^\prime - \psi_2 \right) \left( \tderivative \right) -12 \left( \rho\,\rho^\prime \right)^2\right] \left( \rho^{2} h_{22} \right) \nonumber \\ 
&- \frac14\rho\,\rho^\prime \left[ \left( \rderivative \right) + \left( \tderivative \right) - 6\,\rho\,\rho^\prime \right] \left( \rho^\prime \edt^\prime h_{13} + \rho^\prime \edt\, h_{14} \right) \nonumber \\  
&- \frac14\rho\,\rho^\prime \left[ \left( \rderivative \right) - \left( \tderivative \right) - 6\,\rho\,\rho^\prime \right] \left( \rho\, \edt^\prime h_{23} + \rho\, \edt\, h_{24} \right) \nonumber \\
&- \frac18 \left[ \left( \rderivative \right)^2 - \left( \tderivative \right)^2 - 2\,\rho\,\rho^\prime \left( \wacky \right) \right] \left( 2\,\rho\,\rho^\prime h_{12} \right) \nonumber \\
&+ \frac14 \left[ \left( 5\rho\,\rho^\prime-2\,\psi_2 \right) \left( \rderivative \right) -2\left( 2\,\rho^2\rho^{\prime 2} + \psi_2^2 \right) \right] \left( 2\,\rho\,\rho^\prime h_{12} \right) \nonumber \\
&+ \frac14 \rho\,\rho^\prime \left[ \left(\rderivative \right)^2 - \left( \tderivative\right)^2 - 6\,\rho\,\rho^\prime \left( \rderivative \right) \right] h_{34}.
\end{align}
\end{widetext}
The scalarized versions of $\mathcal{E}_{44}$, $\mathcal{E}_{22}$, and $\mathcal{E}_{24}$ can be obtained by taking a \emph{prime} of the one's with $\mathcal{E}_{33}$, $\mathcal{E}_{11}$ and $\mathcal{E}_{13}$, respectively. The scalarized versions of $\mathcal{E}_{14}$ and $\mathcal{E}_{23}$ can be obtained by taking a \emph{bar} (complex conjugate) of the one's with $\mathcal{E}_{13}$ and $\mathcal{E}_{24}$, respectively. With this, our presentation of the scalarized versions of the 10 components of Einstein tensors is complete. 
\section{Translating GHP to Boyer-Lindquist coordinates}
\label{app:coord}

In this section we translate some of the GHP operators acting on quantities of type (0,0) to Boyer-Lindquist coordinates using the Kinnersley tetrad. 
\begin{align}
\left( \rho\,\kata^\prime + \rho^\prime \kata \right) &= \frac{1}{r}\frac{\partial}{\partial r_{*}} = \frac{\left( 1-\frac{2M}{r} \right) }{r}\frac{\partial}{\partial r}, \\
\left( \rho\,\kata^\prime - \rho^\prime \kata \right) &= \frac{-1}{r}\frac{\partial}{\partial t}, \\
\left( \wacky \right) &= \frac{-\ell(\ell+1)}{r^2},  \\
\left( \bigwacky \right) &= \frac{(\ell-1)(\ell+2)}{-r^2}, \\
\left( \edt\,\edt^\prime+\edt^\prime\edt - 4\,\rho\,\rho^\prime+2\,\psi_2 \right) &= \frac{(\ell-1)((\ell+2)+\frac{6M}{r}}{-r^2}.
\end{align}
When acting on quantities of type $(p,q)$, the $\edt$ and $\edt^\prime$ operators are
\begin{align}
\edt f_{p,q} &= \frac{1}{\sqrt{2}r} \left( \partial_\theta + \frac{i}{\sin\theta}\partial_\phi + \left( \frac{q-p}{2}\right) \cot\theta \right)f_{p,q}, \\
\edt^\prime f_{p,q} &= \frac{1}{\sqrt{2}r} \left( \partial_\theta - \frac{i}{\sin\theta}\partial_\phi + \left(\frac{p-q}{2}\right) \cot\theta \right)f_{p,q}.
\end{align}

Using these, the equation for $I_{\nu E 1}$ is
\begin{align}
&\left( \partial_t^2 - \partial_{r_*}^2 \right) I_{\nu E 1} - \frac{2(r-2\,M)\left[ 36\,M+5(\ell-1)(\ell+2)r \right]}{r^2 \left[ 6\,M+\left( \ell^2+\ell-2 \right) r \right]}\partial_{r_*} I_{\nu E 1} \nonumber \\
&+ \frac{(r-2\,M) \left[ 300\,M^2+30\,M\,r\left( \ell^2+\ell-8 \right) + (\ell-4)(\ell+5)(\ell-1)(\ell+2)r^2\right] }{r^4 \left[ 6\,M+\left( \ell^2+\ell-2 \right) r \right]} I_{\nu E 1} \nonumber \\
=&-\frac{8\,\pi\,(\ell-1)(\ell+2)}{r^3}\partial_{r_*}A^{(0)}_{\ell,m} 
+ \frac{4\sqrt{2}\,\pi\,i\,(\ell-1)(\ell+2)(r-2\,M)}{r^4}\partial_t A^{(1)}_{\ell,m} \nonumber \\
&- \frac{4\,\pi\,(\ell-1)\,\ell\,(\ell+1)(\ell+2)(r-2\,M)^2}{r^6}A_{\ell,m} \nonumber \\
&+ \frac{4\sqrt{2}\,\pi\sqrt{(\ell-1)\,\ell\,(\ell+1)(\ell+2)}(r-2\,M)\left[ 6\,M+\left( \ell^2+\ell-2 \right) r \right]}{r^6}F_{\ell,m}\nonumber \\
&- \frac{4\sqrt{2}\,\pi\,(\ell-1)(\ell+2)\sqrt{\ell\,(\ell+1)}\,(r-2\,M)^2}{r^6}B_{\ell,m} \nonumber \\
&+ \frac{4\,\pi\,(\ell-1)(\ell+2)\left[ 96\,M^2+2\,M\,r\left( 7\,\ell^2+7\,\ell -32\right) + (\ell-1)(\ell+2)\left( \ell^2+\ell-4 \right)r^2 \right]}{r^5\left[ 6\,M+\left( \ell^2+\ell-2 \right) r \right]}A_{\ell,m}^{(0)},
\end{align}
where $A_{\ell,m}^{(0)}$, $A_{\ell,m}^{(1)}$, $A_{\ell,m}$, $B_{\ell,m}$ and $F_{\ell,m}$ are the widely used components of tensor harmonics given in Eq (A1) of \cite{Zerilli_PRD} (where we replaced $L$ with $\ell$), and $I_{\nu E 1}$ is decomposed as sum over ordinary spherical harmonics. 
\bibliography{SWABA}

\end{document}